\documentclass[amsmath,amssymb,twocolumn,prc,nofootinbib,preprint,10pt]{revtex4-2}
\usepackage{bm, mathrsfs}
\usepackage[pdftex]{graphicx, hyperref}

\newcommand{\MeV}{~\mathrm{MeV}}
\newcommand{\fm}{~\mathrm{fm}}
\newcommand{\phiq}{\varphi_{\bm{q}}}
\newcommand{\rLLc}{\texttt{rLL(cut)}}
\newcommand{\rLLd}{\texttt{rLL(delta)}}
\newcommand{\rLLw}{\texttt{rLL(well)}}
\DeclareMathOperator{\erf}{\mathrm{erf}}
\DeclareMathOperator{\sgn}{\mathrm{sgn}}
\newcommand{\Jost}{\mathscr{J}}
\newcommand{\Order}{\mathcal{O}}

\usepackage[pdftex]{color}

\begin{document}
\title{Regularized Lednick\'y-Lyuboshitz formula for higher partial waves in femtoscopy}

\author{Koichi Murase}
\affiliation{Department of Physics, Tokyo Metropolitan University, 1-1
Minami-Osawa, Hachioji, 192-0397, Tokyo, Japan}
\affiliation{RIKEN Interdisciplinary Theoretical and Mathematical Science
Program (iTHEMS), 2-1 Hirosawa, Wako, 351-0198, Saitama, Japan}

\preprint{RIKEN-iTHEMS-Report-25}

\author{Tetsuo Hyodo}
\affiliation{Department of Physics, Tokyo Metropolitan University, 1-1
Minami-Osawa, Hachioji, 192-0397, Tokyo, Japan}
\affiliation{RIKEN Interdisciplinary Theoretical and Mathematical Science
Program (iTHEMS), 2-1 Hirosawa, Wako, 351-0198, Saitama, Japan}

\begin{abstract}
Femtoscopy is one of the promising experimental approaches to put constraints
on interactions between various species of hadrons from the momentum
correlation functions measured in high-energy nuclear collision experiments.
The Koonin--Pratt and Lednick\'y--Lyuboshitz formulae provide useful
expressions of the correlation functions and have been widely used to analyze
the experimental data based on the assumption that the effect of higher partial
waves is negligible.
Those formulae can be generalized for higher partial waves, but the generalized
Lednick\'y--Lyuboshitz formula produces wrong results due to a singular
behavior of the asymptotic wave function at the origin.  In this study, we
attempt to solve the problem by regularizing the generalized
Lednick\'y--Lyuboshitz formula with a cutoff and validate it using the
Koonin--Pratt formula as a reference.  We also show the relationship between
the cutoff in the regularized Lednick\'y--Lyuboshitz and the effective-range
correction in the original Lednick\'y--Lyuboshitz formula.
Using the obtained formula, we investigate the source-size dependence, the
validity of the effective range expansion, and the cutoff dependence of the
correlation function.  We also discuss the interaction dependence using the
heatmap as a function of the scattering-length parameter $1/a_l$ and the
momentum $q$.
\end{abstract}
\maketitle

\section{Introduction}

Femtoscopy in high-energy nuclear collisions has recently attracted attention
as a novel means to experimentally study hadron--hadron interactions using
momentum correlations of hadrons~\cite{Ohnishi:1998at, Morita:2014kza,
ExHIC:2017smd, Fabbietti:2020bfg}.  For femtoscopic studies of hadron--hadron
interactions, systematic measurements of femtoscopic correlation functions of
various hadron combinations have been performed in the ALICE experiment at the
Large Hadron Collider (LHC) in CERN~\cite{ALICE:2018ysd, ALICE:2019hdt,
ALICE:2020mfd, ALICE:2022enj} and the STAR experiment at the Relativistic Heavy
Ion Collider (RHIC) in Brookhaven National Laboratory (BNL)~\cite{STAR:2014dcy,
STAR:2015kha}.  The femtoscopy at lower energy collisions with FAIR, NICA, and
J-PARC-HI is also expected to provide the opportunity to further constrain the
hadron interactions; the strong interaction between hadrons is reported to have
effects on the femtoscopic correlations at lower collision energies in the
HADES experiment at GSI~\cite{Stefaniak:2024fkf}, and the future measurements
at J-PARC-HI are also important~\cite{Hyodo:2025app, Jinno:2024tjh,
Kamiya:2024diw}.

For femtoscopy, the two-particle momentum correlation function is measured in
high-energy collision experiments:
\begin{align}
  C(\bm{q}, \bm{P}) =
  C(\bm{p}_1, \bm{p}_2) = \dfrac{E_1 E_2 (dN_{12}/d\bm{p}_1d\bm{p}_2)}{E_1(dN_1/d\bm{p}_1)\, E_2(dN_2/d\bm{p}_2)},
  \label{def:corr}
\end{align}
where $E_i = E_i(\bm{p}_i)$ is the energy of the particle of species $i$
($i=1,2$) and the momentum $\bm{p}_i$, $dN_i/d\bm{p}_i$ represents the average
number of particles of species $i$ with the momentum $\bm{p}_i$ in an event, and
$dN_{12}/d\bm{p}_1d\bm{p}_2$ represents the average number of pairs of
particles 1 and 2 of the momenta $\bm{p}_1$ and $\bm{p}_2$ in an event.  The
momenta $\bm{p}_1$ and $\bm{p}_2$ are usually represented by the relative and
total momenta $\bm{q}$ and $\bm{P}$.  The average of the correlation function
is often taken over the total momentum $\bm{P}$ after moving to a specific
choice of the inertial frame, such as the pair rest frame ($\bm{P} = 0$).

The correlation function~\eqref{def:corr} is theoretically described by the
Koonin--Pratt (KP) formula~\cite{Koonin:1977fh, Pratt:1984su, Bauer:1992ffu}.
The KP formula is based on the picture in which the observed two particles are
isolated from the rest of the system at emission points, $x_1$ and $x_2$, and
interact with each other as a two-body isolated system until the two-particle
distance becomes sufficiently large.  With an additional approximation that the
relative momentum at emission $\bm{q}^*$ can be replaced by the asymptotic
momentum $\bm{q}$ (which is justified under the smoothness approximation and
the on-shell approximation~\cite{Wiedemann:1999qn, Lisa:2005dd}), the KP
formula gives the two-particle momentum correlation with the integral:
\begin{align}
  C(\bm{q}) &= \int d^3\bm{r} S_{12}(\bm{r}; \bm{q}) |\phiq^{(-)}(\bm{r})|^2,
  \label{eq:KP}
\end{align}
where $\bm{r} = \bm{x}_1 - \bm{x}_2$ is the relative position of the emission
points of the two particles, the relative source function $S_{12}(\bm{r};
\bm{q})$ is the distribution of pairs of the particles 1 and 2 emitted at the
relative position $\bm{r}$ for a given $\bm{q}$, and the wave function
$\phiq^{(-)}(\bm{r})$ represents the amplitude
of finding two interacting particles at the relative distance $\bm{r}$, given
that the final relative momentum of the two particles is $\bm{q}$.  In
practical analyses, the $\bm{q}$-dependence of $S_{12}(\bm{r};\bm{q})$ is
assumed to be negligible: $S_{12}(\bm{r};\bm{q}) = S_{12}(\bm{r})$.
Traditional
femtoscopy is used to measure the size of the emission source through
$S_{12}(\bm{r})$~\cite{Goldhaber:1960sf, HanburyBrown:1956bqd} by assuming
$\phiq^{(-)}(\bm{r})$ to be a plane wave with quantum correlation of identical
particles.  However, the wave function may be modified by the interaction
between the two particles, which can be utilized to constrain the interaction.
The wave function $\phiq^{(-)}(\bm{r})$ is determined by solving the
Schr\"odinger equation (in the nonrelativistic case) under the boundary
condition that the outgoing wave matches that of the plane wave at large
$|\bm{r}|$.  This is comparable to the scattering wave function $\phiq(\bm{r})$
in the scattering theory, which is determined under the boundary condition that
the incoming wave matches that of the plane wave.
The wave function $\phiq^{(-)}(\bm{r})$ is related to the scattering wave
function $\phiq(\bm{r})$ through the complex conjugate, $\phiq^{(-)}(\bm{r}) =
[\phiq(-\bm{r})]^*$, so
$|\phiq^{(-)}(\bm{r})|^2$ in the KP formula~\eqref{eq:KP} is interchangeable
with $|\phiq(\bm{r})|^2$. In the following discussions, we exclusively use
$\phiq(\bm{r})$ for a better analogy with the terminology in the traditional
scattering theory.  We also do not consider the symmetrization and
antisymmetrization for the identical bosons and fermions in the present study.

To extract the information of the interaction from the correlation function
based on the KP formula~\eqref{eq:KP}, one way is to assume a potential between
the particles, solve the Schr\"odinger equation, and compare the result with
experimental data, using a framework such as Correlation Analysis Tool using
the Schr\"odinger equation (CATS)~\cite{Mihaylov:2018rva}.  Another way is to
try to extract physical observables, the scattering length and the effective
range, in a model-independent manner by fitting the Lednick\'y--Lyuboshitz
formula~\cite{Lednicky:1981su} to data.  However, nontrivial assumptions lie
behind the KP and LL formulae~\cite{Wiedemann:1999qn, Lisa:2005dd}.  For
example, the issue of the universality of the source function and the off-shell
ambiguities of the potential~\cite{Epelbaum:2025aan, Gobel:2025afq,
Molina:2025lzw} is an important topic discussed recently.

In our previous study~\cite{Murase:2024ssm}, we focused on the assumption in
the traditional LL formula that only the $s$-wave interaction affects the wave
function in the KP formula, so that the effects of higher partial waves would
be negligible.  In realistic situations, the correlation function receives
corrections from any partial waves in principle.  In fact, peaks associated
with resonances of higher partial waves are seen in the correlation
functions measured in experiments~\cite{ALICE:2019gcn, ALICE:2023wjz}.
We demonstrated cases of potentials for which the contributions of higher
partial waves are nonnegiligible. Assuming the spherical Gaussian source and an
assumption consistent with the traditional LL formula, we also obtained a
generalized version of the LL formula~\cite{Murase:2024ssm}.  However, we found
that the generalized LL formula exhibits a wrong behavior related to an
unphysical property of the asymptotic wave function assumed in the LL formula
near the origin $r=0$.

In this study, we propose an improved version of the LL formula for higher
partial waves and investigate its performance.  In Sec.~\ref{sec:gLL}, after
explaining the KP formula with the spherical source and the LL formula for the
$s$ wave, we first summarize the result of the previous
work~\cite{Murase:2024ssm}, i.e., the KP and LL formulae generalized for higher
partial waves, and explain the problem of the generalized LL formula.  In
Sec.~\ref{sec:rLL}, we propose regularization of the generalized LL formula and
discuss its relationship with the effective-range correction of the traditional
LL formula for the $s$ wave.  In Secs.~\ref{sec:rLL-result}
and~\ref{sec:rLL-resonance}, we demonstrate the performance of the regularized
LL formula for potentials without and with resonances using the
spherical-source KP formula as a reference.  We then discuss the interaction
dependence of the correlation function using the obtained regularized LL
formula in Sec.~\ref{sec:ohnishi}.  Section~\ref{sec:conclusion} is devoted to
the conclusion and summary.  We assume $\hbar = c = 1$, and $\hbar c =
197.327\MeV$ throughout this paper.

\section{KP and LL formulae for higher partial waves}
\label{sec:gLL}

In this section, we first review the KP formula for the spherical
source~\cite{Morita:2014kza} and the LL formula~\cite{Lednicky:1981su} for the
$s$ wave ($l=0$).  We also summarize their generalized versions for higher
partial waves ($l\ge1$)~\cite{Murase:2024ssm} and discuss the problem of the
generalized version of the LL formula.

\subsection{Traditional formulae for the $s$ wave}

With the spherical source function $S_{12}(\bm{r}) = S(r=|\bm{r}|)$, assuming
the wave function to be modified from the plane wave only in the $s$ wave, the
KP formula is known to be rewritten in the following
form~\cite{Morita:2014kza}:
\begin{align}
  C(q) &= 1 + \int_0^\infty dr\,4\pi r^2 S(r) [|R_0(r)|^2 - |j_0(qr)|^2].
  \label{eq:sphKP-s}
\end{align}
This is obtained
by substituting the wave function of the following form in the original KP
formula~\eqref{eq:KP}:
\begin{align}
  \phiq(\bm{r}) = e^{iqz} - j_0(qr) + R_0(r),
  \label{eq:phi.s-correction}
\end{align}
where $q = |\bm{q}|$, $r = |\bm{r}|$, and $z$ is the component of $\bm{r}$ in
the direction of $\bm{q}$. The function $j_l(x)$ is the spherical Bessel
function of the first kind.  The first two terms of
Eq.~\eqref{eq:phi.s-correction} represent components of higher partial waves
($l\ge1$) in the plane wave; the $s$-wave component $j_0(qr)$ is subtracted
from the plane wave $e^{iqz}$.  The third term $R_0(r)$ represents the $s$-wave
component of the interacting wave function $\phiq(\bm{r})$.

Another useful formula for the femtoscopy is the LL formula.  To obtain the LL
formula, we assume the Gaussian form of the width $\sqrt{2}R$ for the relative
source function,
\begin{align}
  S(r) = \frac1{(4\pi R^2)^{3/2}} \exp\Bigl(-\frac{r^2}{4R^2}\Bigr).
  \label{eq:gauss-source}
\end{align}
We also assume that the asymptotic form of the wave function $\phiq(\bm{r}) =
e^{iqz} + f_0(q)e^{iqr}/r$ ($r\to\infty$) can be used to describe the wave
function in the entire $r$ range, where $f_0(q)$ is angle-independent
scattering amplitude, i.e., the $s$-wave ($l=0$) amplitude.  Under those
assumptions, the LL formula for the $s$ wave ($l=0$) is obtained as
\begin{align}
  C_\mathrm{LL}(q) &= 1 + \Delta C_\mathrm{LL}(q), \\
  \Delta C_\mathrm{LL}(q) &= \frac{F_3(r_0 / R)}{2R^2}|f_0(q)|^2 \notag \\
    &\quad + \frac{2F_1(2qR)}{\sqrt{\pi}R}\Re f_0(q) -\frac{F_2(2qR)}{R} \Im f_0(q),
  \label{eq:ll.traditional}
\end{align}
where $F_1(x) = \int_0^x dt\exp(t^2 - x^2)/x$, $F_2(x) = (1-e^{-x^2})/x$, and
$F_3(x) = 1 - x/(2\sqrt{\pi})$.  The symbol $r_0$ denotes the effective range
for the $s$-wave interaction, and the term proportional to $r_0$ is called the
effective-range correction.

\subsection{Generalization for higher partial waves}

In our previous study~\cite{Murase:2024ssm}, we presented expressions for the
KP and LL formula generalized for higher partial waves under assumptions
consistent with the traditional KP and LL formulae for the $s$ wave.

For the traditional expression of the KP formula with the spherical
source~\eqref{eq:sphKP-s}, the wave function is assumed to be modified only in
the $s$-wave channel.  However, in general, all partial-wave components in the
wave function are modified by the interaction.  We here consider the full
partial-wave expansion of the wave function:
\begin{align}
  \phiq(r, \theta) &= \sum_{l=0}^\infty (2l+1) i^l R_l(r) P_l(\cos\theta),
  \label{eq:kpx.wavex}
\end{align}
where $P_l(\cos\theta)$ is the Legendre polynomial of order $l$, and $R_l(r)$
is the wave function for the angular momentum $l$.  The scattering angle
$\theta$ is the angle between the vectors $\bm{q}$ and $\bm{r}$. Substituting
this form of the wave function into the original KP formula~\eqref{eq:KP} and
considering the difference from the plane-wave case [$R_l(r) = j_l(qr)$], we
arrive at the spherical-source KP formula generalized for higher partial
waves~\cite{Murase:2024ssm}:
\begin{align}
  C(q) &= 1 + \sum_{l=0}^\infty \Delta C_l(q),
  \label{eq:sphKP}
\end{align}
where the contribution from the partial wave of the angular momentum $l$ reads
\begin{align}
  \Delta C_l(q) = (2l+1)
  \int_0^\infty dr\,4\pi r^2 S(r) [|R_l(r)|^2 - |j_l(qr)|^2].
  \label{eq:sphKP-partial}
\end{align}
We point out that the deviation of the correlation function from unity is
described by the sum of the contribution from each partial wave because of the
assumption of the spherical source as discussed in Ref.~\cite{Murase:2024ssm}.

The LL formula~\eqref{eq:ll.traditional} can also be generalized for higher
partial waves.  Approximating the wave function of the entire $r$ region with
the asymptotic form in the limit $r\to\infty$ up to $\mathcal{O}(1/r)$,
\begin{align}
  R_l(r) &\approx \frac{1}{qr} \sin\Bigl(qr + \delta_l(q) - \frac{l\pi}2\Bigr),
  \label{eq:asympLL} \\
  j_l(r) &\approx \frac{1}{qr} \sin\Bigl(qr - \frac{l\pi}2\Bigr),
\end{align}
where $\delta_l(q)$ is the phase shift of the angular momentum $l$,
the correlation function~\eqref{eq:sphKP} and~\eqref{eq:sphKP-partial} can be
calculated~\cite{Murase:2024ssm} as
\begin{align}
  C(q)
  &= 1 + \sum_{l=0}^\infty \frac{4\pi (2l+1)!(-1)^l}q \Im[f_l(q) \hat S(q)] \notag \\
  &= 1 + \frac{4\pi}q \Im[f(q,\theta=\pi) \hat S(q)],
  \label{eq:gLL}
\end{align}
where $f_l(q,\theta)$ is the scattering amplitude, and $f_l(q)$ is the
amplitude of the partial wave of the angular momentum $l$,
\begin{align}
  f(q,\theta) = \sum_{l=0}^\infty (2l+1) f_l(q) P_l(\cos\theta).
\end{align}
The function $\hat S$ denotes the Fourier--Laplace transform of the source
function:
\begin{align}
  \hat S(q) = \int_0^\infty dr S(r) e^{2iqr}.
\end{align}


The original LL formula for the $s$ wave ($l=0$) is reproduced except for the
term of the effective-range correction.  Noticing the optical theorem for the
$s$-wave amplitude, $(1/q)\Im f_0(q) = |f_0(q)|^2$, one can rewrite
Eq.~\eqref{eq:ll.traditional} into the following form:
\begin{multline}
  \Delta C_\mathrm{LL}(q) = \biggl[\frac{e^{-(2qR)^2}}{2qR^2} - \frac{r_0}{4\sqrt{\pi} qR^3}\biggr] \Im f_0(q) \\
    + \frac{2F_1(2qR)}{\sqrt{\pi}R}\Re f_0(q).
\end{multline}
Using the Fourier--Laplace transform of the Gaussian source
function~\eqref{eq:gauss-source}:
\begin{align}
  \hat S(q) = \frac{e^{-(2qR)^2}}{(4\pi)^{3/2} R^2} \biggl(\sqrt{\pi} + 2i\int_0^{2qR} dt e^{t^2}\biggr),
  \label{eq:gaussian-FourierLaplace}
\end{align}
one can confirm that this is consistent with the $s$-wave part of the
generalized LL formula~\eqref{eq:gLL}, except that the term proportional to the
effective range $r_0$,
\begin{align}
  \Delta C_\mathrm{LL,eff}(q) = - \frac{r_0}{4\sqrt{\pi} qR^3} \Im f_0(q),
  \label{def:DeltaC-ERC}
\end{align}
is missing in the generalized LL formula~\eqref{eq:gLL}.


However, as reported in Ref.~\cite{Murase:2024ssm}, for higher partial waves $l
\ge 1$, the naive generalization of the LL formula~\eqref{eq:gLL} produces
unphysical behavior due to the singularity of the wave function $R_l(r)$ at the
origin ($r=0$); a naive extrapolation of the asymptotic form for
$r\to\infty$~\eqref{eq:asympLL} to the entire $r$ range causes a singular
behavior at $r\to 0$, $R_l(r) \sim \sin(qr + \delta_l(q) - l\pi/2)/r \propto
1/r$, when the phase shift $\delta_l(q)$ is nonvanishing (but reasonably small
$|\delta_l(q)| < \pi/2$) or, for odd $l$, even when $\delta_l(q)$ vanishes.
This is incompatible with the expectation that $R_l(r)$ is finite at the origin
unless an extreme potential (such as the delta function at the origin $r=0$) is
considered.  The problem becomes significant with odd $l$, for which the
singularity is present even with vanishing phase shift $\delta_l$.  As a
result, the generalized LL formula~\eqref{eq:gLL} gives the opposite sign of
the odd-$l$ contributions to the correlation function compared to the exact
integration based on the KP formula~\cite{Murase:2024ssm}.


To mitigate the unphysical behavior of the naive generalization of the LL
formula, we may consider improving the behavior of the wave function near the
origin.  A similar attempt has actually been performed for the LL formula for
the $s$-wave case by Albaladejo et al.~\cite{Albaladejo:2024lam}.  Actually,
the singularity of the wave function at the origin is present even for the
$s$-wave case ($l=0$), though the singularity $|\varphi|^2 \sim |1/r|^2$ is
canceled with the Jacobian $4\pi r^2$ of the spatial integral, and the LL
formula has provided a good approximation for the $s$-wave contribution when
$R$ is sufficiently large, as demonstrated in existing analyses.  However, even
for the $s$-wave case, the singularity becomes a problem when $R$ is
small~\cite{Albaladejo:2024lam}.  This problem was addressed by using a form of
the asymptotic function regularized at the origin $r=0$, obtained by an
ultraviolet regulator in the half off-shell
$T$-matrix~\cite{Albaladejo:2024lam}.

In this study, we attempt to consider a simpler form of regularization of
partial wave functions $R_l(r)$ for an arbitrary $l$ to obtain a practically useful formula.
A realistic behavior of $R_l(r)$ near the origin is to scale as $R_l(r) \propto
r^l$, due to the strong repulsion by the centrifugal force.
A problem with the naive asymptotic form~\eqref{eq:asympLL} is that it is
missing the effect of the centrifugal force.  Then, one might consider using
the asymptotic form in terms of the spherical Bessel function, which takes
account of the effect of the centrifugal force:
\begin{align}
  R_l(r)
  &= e^{i\delta_l(q)} [\cos\delta_l(q) j_l(qr) - \sin\delta_l(q) y_l(qr)] \notag \\
  &= j_l(qr) + iqf_l(q) h_l(qr).
\end{align}
where $j_l(x)$ and $y_l(x)$ are the spherical Bessel functions of the first and
second kinds, and $h_l(x) = j_l(x) + iy_l(x)$ is the spherical Hankel function
of the first kind.  However, this results in an even worse situation: for
nonvanishing $\delta_l$ with $l\ge1$, the integral of the correlation function
at small $r$ does not converge due to the strong singularity of $y_l(x) \sim
-(2l-1)!!/x^{l+1}$.


Another interest is the extension of the effective-range correction, which is
present in the original LL formula but missing in the generalized LL
formula~\eqref{eq:gLL}.  In the original LL formula, the effective-range
correction has been shown to be effective in improving the description of the
correlation function with the intermediate source sizes $R$, where the
approximation of the wave function by the asymptotic form is expected to become
worse.  For the $s$ wave, we may manually add to Eq.~\eqref{eq:gLL} the
effective-range correction known in the original LL formula.  However, the
expression of the effective-range correction for higher partial waves remains
nontrivial.

\section{LL formula with regularization}
\label{sec:rLL}

To avoid the divergences of the wave function at the origin in the generalized
LL formula~\eqref{eq:gLL}, we here consider employing a different form of the
wave functions, which is regular.  In deriving the original LL formula, we
extrapolate the wave function within the interaction range using the asymptotic
form of the wave function at $r\to\infty$, but there is no strong reason for
using the form at $r\to\infty$ in the entire $r$ range.  We know that the
original LL formula describes the correlation function well, even when the
source size is relatively small.  This implies that replacing a real wave
function with the asymptotic one does not affect the correlation function much
in the case of the $s$-wave interaction.  Therefore, it is reasonable to think
about employing a specific form of a regular wave function as the form of the
extrapolated wave functions.

To determine a regular wave function, we utilize the phase shift $\delta_l(q)$
as an input.  This way of determining the wave function using on-shell
quantities is in line with the original idea of the LL formula for the $s$
wave; in the original LL formula, the correlation function is written in terms
of $f_0(q) = e^{i\delta_0(q)} \sin \delta_0(q) /q$ and mathematical functions.
Then, by substituting the effective range expansion of the $s$-wave amplitude
$f_0(q)$, we may derive the fitting formula for the correlation function with
the $s$-wave interaction.  To follow the same procedure, it is useful to give a
formula to approximately calculate the partial-wave contribution $\Delta
C_l(q)$ using the scattering amplitude $f_l(q)$ [or, equivalently, the phase
shift $\delta_l(q)$], where we can substitute the effective range expansion,
\begin{align}
  f_l(q | a_l, r_l, \ldots) = \frac{q^{2l}}{-1/a_l + r_l q^2/2 + \cdots - iq^{2l+1}}.
  \label{eq:ERE}
\end{align}
To obtain such a correlation formula in terms of $f_l(q)$, we may model a form
of the wave function for a given $q$ and corresponding $f_l(q)$.

In this study, we combine two asymptotic forms at $r\to 0$ and $r\to
\infty$. More specifically, we introduce a cutoff range $r = r_c$, where two
asymptotic forms are switched:
\begin{align}
  R_l(r) &= \begin{cases}
    A j_l(\kappa r), & (r < r_c), \\
    j_l(qr) + iqf_l(q) h_l(qr), & (r > r_c).
  \end{cases}
\end{align}
The cutoff range $r_c$ may be roughly associated with an interaction range
because the wave function at $r > r_c$ is described by the asymptotic form in a
space where the potential is negligible and only the centrifugal force is
effective.
Here, $A$ and $\kappa$ are parameters of the asymptotic form at $r\to 0$, which
should be specified as functions of $\delta_l(q)$ or $f_l(q)$.  Depending on
the way to determine $(A, \kappa)$, we introduce regularized LL formulae with
three different regularizations named \rLLc{}, \rLLd{}, and \rLLw{}.

\subsection{\rLLc: Regularized LL formula with cutoff}

The simplest choice of $(A, \kappa)$ is to use the asymptotic form of the plane
wave, i.e., $A = 1$ and $\kappa = q$ independent of $\delta_l$.  In this case,
the partial-wave contributions to the correlation function become
\begin{align}
  \Delta C_l(q) &= - \frac{4\pi(2l+1)}q \Im[f_l \hat S_l(q;r_c)],
    \label{eq:correction.hatS} \\
  \hat S_l(q;r_c) &= \int_{r_c}^\infty dr S(r) \hat h_l^2(qr),
    \label{def:Sc}
\end{align}
where $\hat h_l(x) = r h_l(x)$ is the Riccati--Hankel function of the first
kind.  An important difference from the naive LL formula~\eqref{eq:gLL} is the
integral domain of the Fourier--Laplace transform.  The integration in the
domain $[0, r_c]$ vanishes due to the cancellation of the wave function,
$|R_l(r)|^2 - |j_l(qr)|^2 = 0$, in the integrand of the spherical-source KP
formula~\eqref{eq:sphKP-partial}.  In the generalized LL formula, the integral
domain is the entire range of $r > 0$, which caused the divergence of the
integral at $r=0$.  In this new formula, the divergence is regularized by
effectively introducing a cutoff $r > r_c$ in the integration.  Another
difference is that the full spherical Hankel function $\hat h_l^2(x)$ is used
as the integration kernel, instead of $e^{2ix}\; [\sim (-1)^{l+1}\hat h_l^2(x)
+ \mathcal{O}(1/x)]$.  Thus, $\hat S_l$ now depends on the angular momentum
$l$.  We call this version of the regularized LL formula \rLLc{}.

It should be noted that there is no corresponding physical potential that
reproduces the wave function of \rLLc{} with the parameters $(A, \kappa) = (1,
q)$ because the wave function is discontinuous at $r=r_c$ with these
parameters.  Thus, we also consider formulae with more realistic wave functions
and compare the obtained formulae.

\subsection{\rLLd: Regularized LL formula with continuous wave function}
\label{sec:rLLd}

Another choice of the parameters $(A, \kappa)$ is to fix $\kappa=q$ and choose
$A$ so that the wave function is continuous at $r = r_c$.  The continuous
condition is explicitly written as $A j_l(qr_c) = j_l(qr_c) + iqf_l(q)h_l(qr_c)$, but
in obtaining the final formula, it is easier to use the fact that the integrand
in the spherical-source KP formula~\eqref{eq:sphKP-partial} is continuous at $r=r_c$:
\begin{align}
  (|A|^2-1)j_l^2(qr_c)
  &= |j_l(qr_c) + iqf_l(q)h_l(qr_c)|^2 - j_l^2(qr_c) \notag \\
  &= -q\Im[f_l(q)h_l^2(qr_c)].
\end{align}
To obtain the second equality, we used the optical theorem for partial waves,
$|f_l(q)|^2 = (1/q)\Im f_l(q)$.  The resulting modification of the correlation
function is
obtained in the same form as Eq.~\eqref{eq:correction.hatS}, but with a
different expression of $\hat S_l$:
\begin{align}
  \hat S_l(q;r_c) &= \frac{h_l^2(qr_c)}{j_l^2(qr_c)} \int_0^{r_c} dr S(r)\hat j_l^2(qr)
    + \int_{r_c}^\infty dr S(r) \hat h_l^2(qr).
  \label{def:Sd}
\end{align}
The second term is equivalent to Eq.~\eqref{def:Sc}, but the first term is a
new contribution.  We call this version of the regularized LL formula
\texttt{rLL(delta)}.  This setup corresponds to the wave function with the
delta-shell potential at $r = r_c$, $V(r) = V_0\delta(r-r_c)$, with
\begin{multline}
  V_0 = \frac{q}{2\mu} \partial_x \{\ln[\cos\delta_l(q) j_l(x) \\
  - \sin\delta_l(q) y_l(x)] - \ln j_l(x)\}|_{x=qr_c}.
  \label{eq:V0d}
\end{multline}
The above expression of $V_0$ depends on $q$ and $l$ in general so that it
reproduces the $q$ dependence of a given $\delta_l(q)$.  When the actual
potential giving $\delta_l(q)$ is the delta-shell potential at $r = r_c$,
Eq.~\eqref{eq:V0d} becomes independent of $q$, and \rLLd{} reproduces the exact
KP formula for the delta-shell potential.

\subsection{\rLLw: Regularized LL formula with potential well}
\label{sec:rLLw}

\begin{figure}[thb]
  \centering
  \includegraphics[width = 0.45\textwidth]{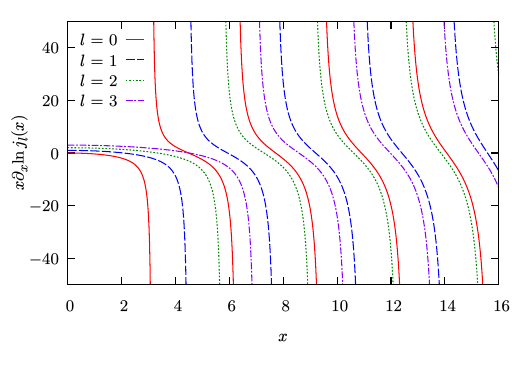}
  \caption{$y = x\partial_x \ln j_l(x)$ for different angular-momentum quantum
  numbers $l$.  The solid red, blue dashed, green dotted, and purple
  dash-dotted lines correspond to $l = 0$, $1$, $2$, and $3$, respectively.}
  \label{fig:xdlnj}
\end{figure}

Yet another way of determining the parameters $(A, \kappa)$ is to require both
$R_l(r)$ and its first-order derivative $\partial R_l(r)/\partial r$ to be
continuous at $r = r_c$.  This corresponds to the wave function of the
square-well potential of the height,
\begin{align}
  V_0 &= \frac{q^2 - \kappa^2}{2\mu},
  \label{eq:V0w}
\end{align}
and width
$r_c$.  The wave number $\kappa$ is given by $\kappa = x/r_c$, where $x$ is
determined by solving the equation,
\begin{align}
  x \partial_x \ln j_l(x) &=
    \frac{[j_l'(qr_c)\cos\delta_l - y_l'(qr_c)\sin\delta_l] qr_c
      }{j_l(qr_c)\cos\delta_l - y_l(qr_c)\sin\delta_l},
  \label{eq:rLLw-solution}
\end{align}
where $j_l'(x) = \partial_x j_l(x)$ and $y_l'(x) =\partial_x y_l(x)$ are the
derivatives of the spherical Bessel functions of the first and second kinds.
The right-hand side can be explicitly evaluated for given $q$, $r_c$, and
$\delta_l$.  Thus, $x$ can be determined by finding the intersection of the
curve $y=x\partial_x \ln j_l(x)$ and the horizontal line whose level is
specified by the right-hand side of Eq.~\eqref{eq:rLLw-solution}.
Figure~\ref{fig:xdlnj} shows $y=x\partial_x \ln j_l(x)$ for different angular
momenta $l$.  We see that the curve sweeps from $\infty$ to $-\infty$ between
every pair of two adjacent zeroes of $j_l(x)$, so there are an infinite number
of positive solutions $x\,(\ge0)$ in general, where we choose the solution
between the first and second zeroes of $j_l(x)$ in the present
study~\footnote{We might consider a better way to pick up the ``best'' solution
from the infinite number of solutions, but we do not bother to invent a
complicated prescription. We would rather keep the method simple and clear,
although this specific choice of the solution might cause a glitch.}.  Once
$\kappa$ is determined, the solution for $A$ is given by
\begin{align}
  A
  &= \frac{j_l(qr_c) + iqf_l(q) h_l(qr_c)}{j_l(\kappa r_c)} \notag \\
  &= e^{i\delta_l}\frac{\cos\delta_l\, j_l(qr_c) - \sin\delta_l\, y_l(qr_c)}{j_l(\kappa r_c)}.
\end{align}
The partial-wave contribution to the correlation function is obtained as
\begin{align}
  &\Delta C_l(q)
  = -\frac{4\pi(2l+1)}q \Im[f_l \hat S_l(q;r_c)] \notag \\
  &\qquad + \frac{4\pi(2l+1)}{q^2}\biggl[\frac{j_l^2(qr_c)}{j_l^2(\kappa r_c)} - 1\biggr]
    \int_0^{r_c} dr\,S(r) \hat j_l^2(\kappa r),
  \label{eq:correction.hatSw}
\end{align}
with
\begin{align}
  \hat S_l(q;r_c) &= \frac{h_l^2(qr_c)}{j_l^2(\kappa r_c)} \int_0^{r_c} dr S(r)\hat j_l^2(\kappa r)
    + \int_{r_c}^\infty dr S(r) \hat h_l^2(qr).
  \label{def:Sw}
\end{align}
We call this version of the regularized LL formula \texttt{rLL(well)}.

In the same way as \rLLd{} reproduces the KP formula for the delta-shell
potential at $r_c$, \rLLw{} would reproduce the one for the square-well
potential of the width $r_c$.  When $\delta_l(q)$ is given by the square-well
potential, the potential height determined by Eq.~\eqref{eq:V0w} becomes
independent of $q$ and matches the actual potential height, as long as the
proper solution to Eq.~\eqref{eq:rLLw-solution} is chosen.  As a result,
\rLLw{} reproduces the KP formula exactly.

\subsection{Relationship with effective-range correction}

Finally, we show that the effective-range correction in the original LL formula
for the $s$ wave can be related to the effect of the regularization by \rLLc{}.

The generalized LL formula (without the effective-range correction) for the $s$
wave ($l=0$) can be obtained by taking the limit of $r_c\to 0$ in any of the
regularized LL formulae, \rLLc{}, \rLLd{}, and \rLLw{}~\footnote{For higher
partial waves, it is insufficient to take the limit $r_c\to 0$ because the
asymptotic form of the wave function also needs to be switched from the
spherical Bessel functions to the simple $(1/qr)\sin(qr+\delta_l(q)-l\pi/2)$.}.
For the present purpose, we focus on \rLLc{}.  The effect of the regularization
in \rLLc{} is evaluated as the difference of the $s$-wave contribution $\Delta
C_0$ between the regularized LL formula (with the finite $r_c$) and the
generalized LL formula (with $r_c\to0$):
\begin{align}
  \Delta C_{0,\mathrm{cutoff}}
  &:= \Delta C_0(q;r_c) - \Delta C_0(q;0) \notag \\
  &= \frac{4\pi}{q} \Im \biggl[f_0(q)\int_0^{r_c} dr\,S(r) \hat h_0^2(qr)\biggr].
  \label{eq:DeltaC-cutoff}
\end{align}
In the low-momentum region $q \ll 1/r_c$, we may consider the lowest order in
$q$, $\hat h_0^2(qr) = -e^{2iqr} \approx -1 + \mathcal{O}(qr_c)$ ($r<r_c$), to
simplify the integral:
\begin{align}
  \Delta C_{0,\mathrm{cutoff}}
  &= -\frac{4\pi}{q} \biggl[\int_0^{r_c} dr\,S(r)\biggr] \Im f_0(q) [1+\mathcal{O}(qr_c)]\notag \\
  &= -\frac1{2qR^2} \erf\Bigl(\frac{r_c}{2R}\Bigr) \Im f_0(q) [1+\mathcal{O}(qr_c)],
\end{align}
where $\erf(x) = (2/\sqrt{\pi}) \int_0^x e^{-t^2} dt$ is the Gauss error
function.  If the source size is sufficiently larger than the typical
interaction range, $R \gg r_c$, we may approximate $\erf(r_c/2R) =
(2/\sqrt{\pi})(r_c/2R) + \mathcal{O}\bigl((r_c/2R)^3\bigr) \approx
r_c/\sqrt{\pi}R$.  With this approximation, the correction by the introduction
of the cutoff $r_c$ is obtained as
\begin{align}
  \Delta C_{0,\mathrm{cutoff}}
  &\approx -\frac{r_c}{2\sqrt{\pi}qR^3} \Im f_0(q) \notag \\
  &\quad \times \Bigl[1+\mathcal{O}(qr_c) + \mathcal{O}\Bigl(\frac{r_c^2}{R^2}\Bigr)\Bigr].
\end{align}
This structure is the same as the effective-range
correction~\eqref{def:DeltaC-ERC} in the traditional LL formula. We identify a
simple relation $r_c = r_0/2$ by equating both
expressions~\eqref{def:DeltaC-ERC} and~\eqref{eq:DeltaC-cutoff} while ignoring
higher-order terms in $qr_c$ and $r_c/R$.

This implies that the regularization of the LL formula has a similar effect as
the effective-range correction in the conventional LL formula for the $s$ wave.
In fact, both contributions, $\Delta C_\mathrm{LL,eff}$ and $\Delta
C_{0,\mathrm{cutoff}}$, aim at taking account of the correction of the wave
function in the interaction region from the asymptotic form.
In this sense, the regularized LL formula $\Delta C_l$ may be regarded as one
way to extend the effective-range correction in the traditional LL formula
($l=0$) to the cases with higher partial waves ($l\ge1$).

\section{Results with Gaussian potentials}
\label{sec:rLL-result}

In this section, assuming model potentials with realistic interaction ranges
and strengths, we examine the regularized LL formulae, \rLLc{}
[Eqs.~\eqref{eq:correction.hatS} and~\eqref{def:Sc}], \rLLd{}
[Eqs.~\eqref{eq:correction.hatS} and~\eqref{def:Sd}], and \rLLw{}
[Eqs.~\eqref{eq:correction.hatSw} and~\eqref{def:Sw} with a solution $x=\kappa
r_c$ to Eq.~\eqref{eq:rLLw-solution}], for higher partial waves.  We use the
spherical-source KP formula~\eqref{eq:sphKP-partial} as a reference for the
expected behavior of the correction of the correlation function by higher
partial waves.

To obtain the wave function $R_l(r)$ and determine the corresponding phase
shift $\delta_l(q)$, and the partial-wave scattering amplitude $f_l(q) =
e^{i\delta_l(q)}\sin \delta_l(q)/q$, we solve the Schr\"odinger equation,
\begin{align}
  \biggl[
    -\frac{\partial^2}{\partial r^2} - \frac2{r}\frac{\partial}{\partial r}
    + \frac{l(l+1)}{r^2}
    + 2\mu V(r)
  \biggr] R_l(r) &= q^2 R_l(r),
  \label{eq:kpx.schro}
\end{align}
for each angular momentum $l$ under a given short-range local potential $V(r)$.  The
details of the numerical method is described in
Appendix~\ref{app:numerics}. Motivated by the two-nucleon system, we assume the
reduced mass $\mu = 469\MeV\; (\sim m_N/2)$ in this study.

For testing purposes, we introduce several variations for the potential $V(r)$.
Although it would be desirable if we could use realistic potentials of specific
hadron pairs as examples, realistic potentials for higher partial waves have
not been well established because the data is limited for $l \gtrsim 3$.  It is
noted that the local potential in general depends on $l$ phenomenologically.
Here, we instead introduce three variations of $l$-independent Gaussian central
potentials:
\begin{align}
  V_\alpha(r) &= \sum_{k=1}^{N_\alpha} A_{\alpha k} e^{-B_{\alpha k} r^2},
  \quad (\alpha = \mathrm{2G}, \mathrm{1GA}, \mathrm{1GR}),
  \label{def:V}
\end{align}
where $\alpha$ is the label of the potential, $N_\alpha$ is the number of
Gaussian forms in potential $\alpha$, and $A_{\alpha k}$ and $B_{\alpha k}$ are
the magnitude and the exponent of the $k$th Gaussian form in potential
$\alpha$.  The parameter values for the potentials are summarized in
Table~\ref{tab:gaussian-potential-params}\@.  The potential $V_\mathrm{2G}(r)$
is a two-range Gaussian potential, whose parameters are motivated by the
spin-triplet part of the Minnesota potential~\cite{Thompson:1977zz}.  The
potential $V_\mathrm{1GA}(r)$ is an attractive one-range Gaussian potential
obtained by picking up only the long-range part of $V_\mathrm{2G}(r)$.  The
potential $V_\mathrm{1GR}(r) = -V_\mathrm{1GA}(r)$ is a repulsive version of
$V_\mathrm{1GA}(r)$.  The two-range Gaussian potential $V_\mathrm{2G}(r)$ has a
shallow bound state in the $s$ wave ($l=0$) with the binding energy
$2.193\MeV$, which corresponds to the deuteron.  The attractive one-range
Gaussian potential $V_\mathrm{1GA}(r)$ has a bound state in the $s$ wave with
the binding energy $26.87\MeV$.  There are no significant resonances with these
potentials.


\begin{table}[thb]
  \centering
  \begin{tabular}{cccccc}
  \hline\hline
  $\alpha$ & $N_\alpha$ & $k$ & $A_{\alpha k}~[\mathrm{MeV}]$ & $B_{\alpha k}~[\mathrm{fm}^{-2}]$ & $B^{-1/2}_{\alpha k}~[\mathrm{fm}]$ \\ \hline
  2G  & 2 & 1 & $200$   & $1.487$ & $0.820$ \\
      &   & 2 & $-178$  & $0.639$ & $1.25$  \\
  1GA & 1 & 1 & $-178$  & $0.639$ & $1.25$  \\
  1GR & 1 & 1 & $178$   & $0.639$ & $1.25$ \\ \hline\hline
  \end{tabular}
  \caption{Potential parameters for multi-range Gaussian central
  potentials~\eqref{def:V}. The symbol $N_\alpha$ represents the number of
  Gaussian functions of potential $\alpha$, and $A_{\alpha k}$ and $B_{\alpha
  k}$ specify the parameters of the $k$th Gaussian in potential $\alpha$. The
  last column shows the potential range corresponding to $B_{\alpha k}$.}
  \label{tab:gaussian-potential-params}
\end{table}

\begin{figure}[thb]
  \centering
  \includegraphics[width = 0.45\textwidth]{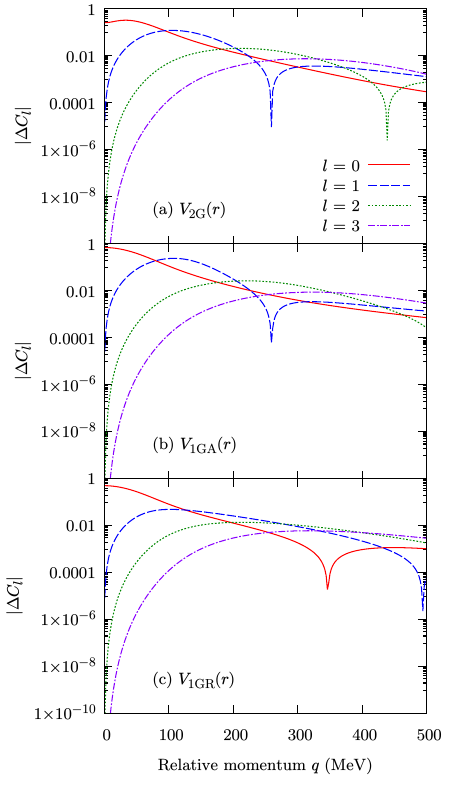}
  \caption{Magnitudes of partial-wave contributions to the correlation
  functions.  Panels (a)--(c) correspond to different potentials,
  $V_\mathrm{2G}(r)$, $V_\mathrm{1GA}(r)$, and $V_\mathrm{1GR}(r)$,
  respectively. The line styles for different $l$'s are the same as
  Fig.~\ref{fig:xdlnj}. The source size is fixed to $R = 2.0\fm$, and the
  reduced mass is $\mu = 469\MeV$.}
  \label{fig:kp-higher}
\end{figure}


\begin{figure}[thb]
  \centering
  \includegraphics[width = 0.47\textwidth]{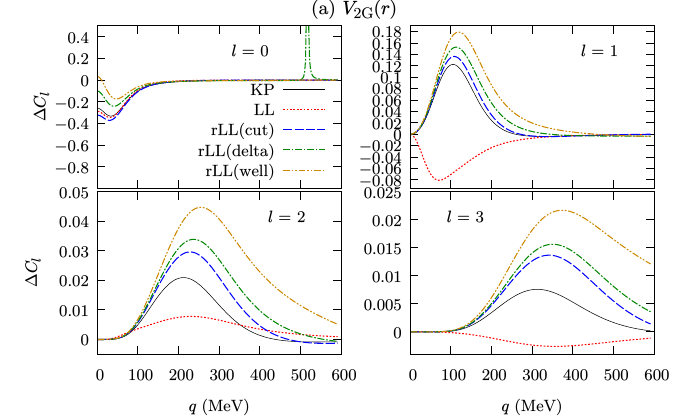}\\
  \includegraphics[width = 0.47\textwidth]{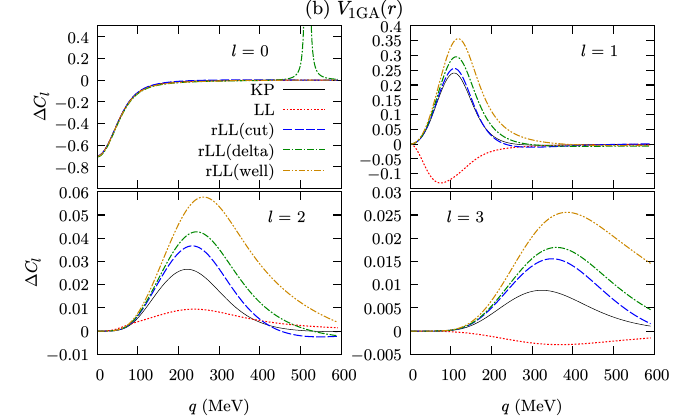}\\
  \includegraphics[width = 0.47\textwidth]{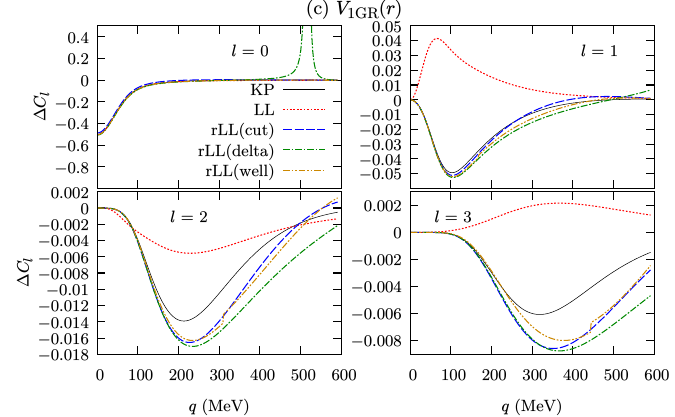}
  \caption{Corrections to correlation functions with different asymptotic forms
  (rLL) with different potentials. The reduced mass is $\mu = 469\MeV$, the
  source size is $R = 2\fm$, and the cutoff is $r_c=1.2\fm$.}
  \label{fig:rLL-compare}
\end{figure}

\begin{figure}[thb]
  \centering
  \includegraphics[width = 0.235\textwidth]{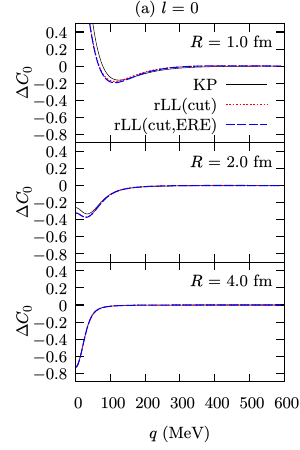}
  \includegraphics[width = 0.235\textwidth]{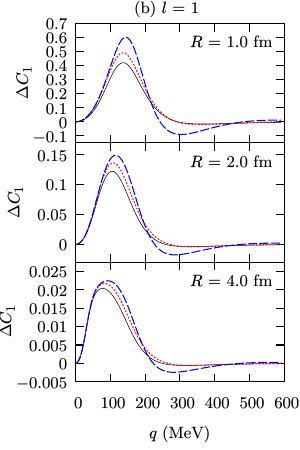}\\[0.5em]
  \includegraphics[width = 0.235\textwidth]{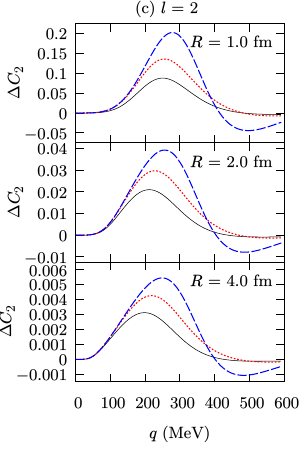}
  \includegraphics[width = 0.235\textwidth]{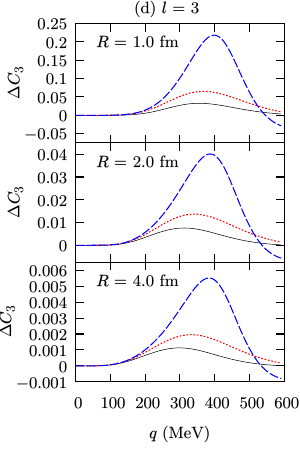}
  \caption{Source-size $R$ dependence of partial-wave contributions to the
  correlation function. We also compare the results with and without the
  effective range expansion. The solid black line shows the exact integration
  by the KP formula, and the red dashed lines show the regularized LL formula
  with the exact $f_l(q)$.  The blue dotted lines show the regularized LL
  formula with $f_l(q)$ approximated by the effective range expansion up to
  $\mathcal{O}(q^2)$: $f_l(q) = q^{2l}/(-1/a_l + r_lq^2/2 - iq^{2l+1})$. The
  potential is $V_\mathrm{2G}(r)$, the reduced mass is $\mu = 469\MeV$, and the
  cutoff is $r_c=1.2\fm$.}
  \label{fig:rLL-compare-R}
\end{figure}

\begin{figure}[thb]
  \centering
  \includegraphics[width = 0.49\textwidth]{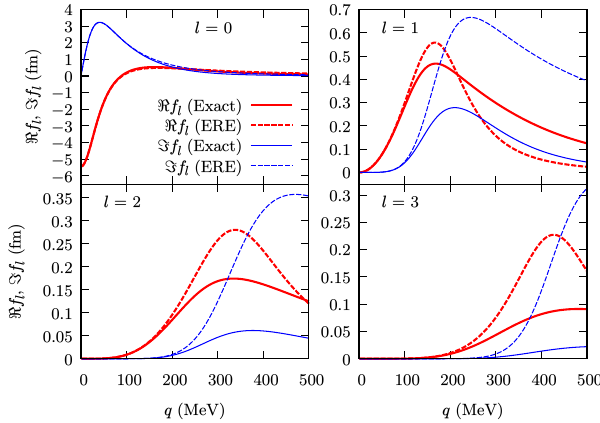}
  \caption{Partial-wave amplitudes $f_l(q)$ as functions of the momentum $q$.
  The solid lines show the exact $f_l(q)$ for the potential, and the dashed
  lines show the effective range expansion up to $\mathcal{O}(q^2)$: $f_l(q) =
  q^{2l}/(-1/a_l + r_lq^2/2 - iq^{2l+1})$.  The bold red lines correspond to
  the real parts of $f_l(q)$, and the thin blue lines to the imaginary parts.
  The potential is $V_\mathrm{2G}$, and the reduced mass is $\mu = 469\MeV$.}
  \label{fig:compare-R-f}
\end{figure}

\begin{figure}[thb]
  \centering
  \includegraphics[width = 0.49\textwidth]{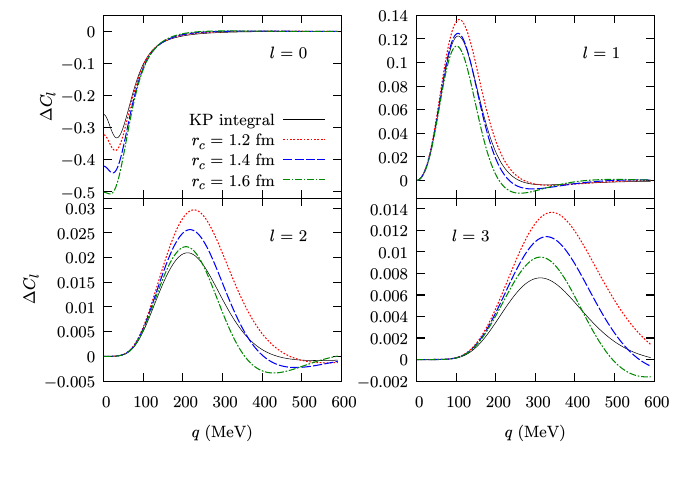}
  \caption{Cutoff $r_c$ dependence of partial-wave contributions to the
  correlation function. The potential is $V_\mathrm{2G}(r)$, the reduced mass
  is $\mu = 469\MeV$, and the source size is $R=2\fm$.}
  \label{fig:rLL-compare-c}
\end{figure}

We first check the behavior of the correction to the correlation function by
higher partial waves using the reference, i.e., the spherical-source KP
formula~\eqref{eq:sphKP}.

In Fig.~\ref{fig:kp-higher}, we compare the magnitudes of the contributions
from different partial waves.  We assumed the source size $R = 2.0\fm$.  The
lower-momentum region is dominated by the $s$-wave ($l=0$) component. However,
looking at a higher-momentum region, we notice that the dominant partial wave
changes in turn.  The momenta at which the dominant partial wave switches
depend on the potential quantitatively, but the qualitative behavior is similar
among the three potentials.  The typical width of the momentum range of each
partial wave is about $100$--$200\MeV$. This scale is comparable to $q \sim l /
r \sim \sqrt{B_i}l \sim 160l\MeV$, considering $l \sim qr$ from the orbital
angular momentum $\bm{L} = \bm{r} \times \bm{q}$.  We note that the sharp
structures in Fig.~\ref{fig:kp-higher} similar to reversed cusps are simply due
to zeroes of the partial-wave contribution $\Delta C_l(q) = 0$ plotted in the
logarithmic scale.  We also find that the peak magnitudes of partial waves
become smaller for larger $l$ in these cases (without significant resonances).

\subsection{Regularization dependence}
\label{sec:rLL-compare-reg}

We first compare the three regularized LL formulae with the (generalized) LL
formula and the exact integration of the KP formula.  For the LL formulae, we
use the exact momentum dependence of the phase shift $\delta_l(q)$ obtained by
solving the Schr\"odinger equation.  The accuracy of the effective range
expansion for the scattering amplitude $f_l(q)$ will be discussed later in
Sec.~\ref{sec:rLL-ere}. Figure~\ref{fig:rLL-compare} shows the partial-wave
contributions for the potentials $V_\mathrm{2G}(r)$, $V_\mathrm{1GA}(r)$, and
$V_\mathrm{1GR}(r)$.

The solid black lines in Fig.~\ref{fig:rLL-compare} show the result of the
exact integration of the KP formula.  For higher partial waves ($l\ge1$), we
see a peak (or a dip) at a finite $q$.  In contrast to the $s$-wave case
($l=0$), this peak is usually not associated with a resonance, but it comes
from a trivial restriction $\Delta C_l(q=0) = 0$ specific to higher partial
waves ($l\ge1$), which will be discussed in Sec.~\ref{sec:ohnishi} in more
detail.  Since the higher-partial-wave contribution vanishes at $q=0$ and
$q\to\infty$, it has at least one maximum or minimum unless the contribution
vanishes at all $q$.

The red dashed lines in Fig.~\ref{fig:rLL-compare} show the results of the
traditional LL formula for $l=0$ with the effective-range correction and the
generalized LL formula for $l\ge1$ without the effective-range correction. As
discussed in Ref.~\cite{Murase:2024ssm}, while the traditional LL formula
($l=0$) is in good agreement with the KP formula, the generalized LL formula
($l\ge1$) fails to describe the qualitative behavior of the KP formula.  In
particular, it exhibits the opposite sign of the result for the odd orders of
the partial waves.

The other lines in Fig.~\ref{fig:rLL-compare} show the regularized LL formulae
with three different regularization methods.  For these lines, the cutoff
$r_c=1.2\fm$ is chosen based on the range $B_{\mathrm{2G},2}^{-1/2} = 1.25\fm$
of the long-range part of $V_\mathrm{2G}(r)$.  Although the three
regularizations give quantitatively different results for $l\ge1$, their
qualitative behaviors match that of the KP formula for all $l$ in most $q$.
While the relative differences from the KP formula become larger for a larger
$l$, the absolute values of the differences are of the same order for all $l$
for each potential.  While \rLLc{} performs the best for the attractive
potentials, $V_\mathrm{2G}(r)$ and $V_\mathrm{1GA}(r)$, \rLLw{} gives similar
or better results for the repulsive potential $V_\mathrm{1GR}(r)$. This implies
that the best regularization method depends on the potential.  In fact, \rLLd{}
and \rLLw{} are expected to give the exact results for the delta-shell
potential at $r = r_c$ and the square-well potential of the width $r_c$,
respectively, as discussed in the last paragraphs of Secs.~\ref{sec:rLLd}
and~\ref{sec:rLLw}.

However, we find pathological behaviors of \rLLd{} and \rLLw{} at specific $q$
values, although they produce consistent results with most $q$ values.  For
example, \rLLd{} for $l=0$ has a divergent peak around $q=500~\mathrm{MeV}$ for
all potentials.  This divergence comes from a zero of $j_l(qr_c)$ in the
denominator of the first term of Eq.~\eqref{def:Sd}.  In general, this
divergence happens at every zero of $j_l(qr_c)$, unless the zero cancels with a
zero in the numerator.  For \rLLw{}, we observe step
structures in Fig.~\ref{fig:rLL-compare}~(c) for $l=2$ and $3$.  This is caused
by the jump of our specific choice of a solution from the infinite number of
solutions to Eq.~\eqref{eq:rLLw-solution}.  Therefore, the regularization
method \rLLc{} is considered to produce simple and stable predictions overall.
In the following discussions, we exclusively consider \rLLc{} as a practical
version of the regularized LL formula.

\subsection{Source-size dependence}
\label{sec:rLL-R}
\label{sec:rLL-ere}

Figure~\ref{fig:rLL-compare-R} shows the source-size dependence of the
partial-wave contributions to the correlation function, $\Delta C_l(q)$, for
the potential $V_\mathrm{2G}(r)$.  The solid black lines show the result from
the exact integration of the KP formula.  The shape of the $s$-wave
contribution changes as $R$ changes, which is related to the existence of a
shallow bound state in the $s$-wave channel mentioned above.  In the other
partial waves without bound states, the shapes of the contributions from higher
partial waves ($l\ge1$) do not depend on $R$ significantly.  The peak positions
become slightly smaller as $R$ increases.  While the shape does not depend on
$R$, the magnitudes of the contributions decrease significantly as $R$
increases, which is consistent with the behavior for the $s$ wave without bound
states.  When $R$ is sufficiently large, the partial-wave contributions become
almost negligible.

The red dashed lines in Fig.~\ref{fig:rLL-compare-R} show the results of
\rLLc{} with the exact amplitudes $f_l(q)$ (or equivalently, the exact phase
shifts $\delta_l(q)$).  The blue dotted lines show the results of \rLLc{} with
the amplitudes approximated by the effective range expansion up to
$\mathcal{O}(q^2)$, $f_l(q) \approx q^{2l}/(-1/a_l + r_lq^2/2 - iq^{2l+1})$.
Both qualitatively reproduce the behavior of the KP formula.  The shape of the
$q$ dependence is similar and does not depend on $R$ significantly.  However,
with the effective range expansion, quantitative differences are sizable for
higher partial waves ($l\ge1$) at large $q$, $q \ge 100$--$200\MeV$.  This is
because the effective range expansion of $f_l(q)$ becomes worse at larger $q$.
Figure~\ref{fig:compare-R-f} shows the real and imaginary parts of $f_l(q)$
with and without the effective range expansion.  The $s$-wave amplitude ($l=0$)
is well described by the effective range expansion up to $\mathcal{O}(q^2)$,
while the effective range expansion of the amplitudes for higher partial waves
($l\ge1$) significantly deviates from the exact one value at $q\gtrsim 100\MeV$
($l=1$), $q\gtrsim 150\MeV$ ($l=2$), and $q\gtrsim 200\MeV$ ($l=3$).  The
momentum where the effective range expansion starts to fail in describing
$f_l(q)$ is consistent with the one for $\Delta C_l(q)$.  Therefore, the large
deviation of the blue dotted lines in Fig.~\eqref{fig:rLL-compare-R} is due to
the quality of the effective range expansion rather than the problem of the
regularized LL formula (with the exact $f_l(q)$), and it may be systematically
improved by considering higher orders in the effective range expansion.
Nevertheless, in low $q$ regions where the effective range expansion works
well, the regularized LL formula also works well.

\subsection{Cutoff dependence}
\label{sec:rLL-cutoff}

In the regularized LL formulae, the cutoff parameter $r_c$ is a free
parameter, and there is no clear way to specify the value externally.  One way
is to fix the cutoff to be a reasonable value of the expected scale in advance.
In such a case, it is important to check how stable the result is with different
choices of the cutoff parameter $r_c$.
Figure~\ref{fig:rLL-compare-c} shows the cutoff dependence of the partial-wave
contributions $\Delta C_l(q)$ by \rLLc{} with the exact $f_l(q)$ for the
potential $V_\mathrm{2G}(r)$.  The solid black line shows the exact integration
of the KP formula.  The red dashed, blue dotted, and green dash-dotted lines
show the results with different regularization cutoffs $r_c = 1.2$, $1.4$, and
$1.6\fm$, respectively.

We observe that the cutoff mainly changes the magnitude of the partial-wave
contribution.  It also slightly shifts the peak position of the structures.  By
adjusting the value of the cutoff $r_c$, the magnitude of the partial-wave
contribution can be made close to the exact integration of the KP formula.  In
the present case of $V_\mathrm{2G}(r)$, $r_c = 1.4$--$1.6\fm$ roughly
reproduces the magnitude of the KP formula.  In particular, \rLLc{} reproduces
the KP formula very well with $r_c=1.4\fm$ in the $p$-wave case ($l=1$). For
the cases of $l=2$ and $3$, the shape of the partial-wave contribution is
different, while $r_c \simeq 1.6\fm$ reproduces the magnitude.  Those values
have the same order as the Gaussian range parameter $B_{\mathrm{2G},2}^{-1/2} =
1.25\fm$ of the long-range part of $V_\mathrm{2G}(r)$, though they do not match
exactly.  For reference, the effective-range parameters $r_l$ for the potential
$V_\mathrm{2G}(r)$ are $r_0 = 1.758\fm$ ($l=0$), $r_1 = 0.05421\fm^{-1} =
(18.45\fm)^{-1}$ ($l=1$), $r_2 = 9.059\fm^{-3} = (0.4797\fm)^{-3}$ ($l=2$), and
$r_3 = 89.45\fm^{-5} = (0.4071\fm)^{-5}$ ($l=3$), which are not necessarily
close to the actual interaction range and do not match the cutoff compared to
the actual range in $V_\mathrm{2G}(r)$.

\section{Results with resonances}
\label{sec:rLL-resonance}


\begin{table}[thb]
  \centering
  \begin{tabular}{cccccc}
  \hline\hline
  $\alpha$ & $N_\alpha$ & $k$ & $A_{\alpha1}~[\mathrm{MeV}]$ & $B_{\alpha k}~[\mathrm{fm}^{-2}]$& $B^{-1/2}_{\alpha k}~[\mathrm{fm}]$ \\ \hline
  $\mathrm{L1}$ & 1 & 1 & 700  & 0.639 & 1.25 \\
  $\mathrm{L2}$ & 1 & 1 & 1500 & 0.639 & 1.25 \\
  $\mathrm{L3}$ & 1 & 1 & 2500 & 0.639 & 1.25 \\ \hline\hline
  \end{tabular}
  \caption{Parameters of the potentials with resonances.  $V_\mathrm{L1}(r)$,
  $V_\mathrm{L2}(r)$, $V_\mathrm{L3}(r)$ have resonances in the angular
  momentum channels $l=1$, $2$, and $3$, respectively.  The meaning of the
  parameters is the same as in Table~\ref{tab:gaussian-potential-params}.}
  \label{tab:reso-potparams}
\end{table}

\begin{figure}[thb]
  \centering
  \includegraphics[width = 0.45\textwidth]{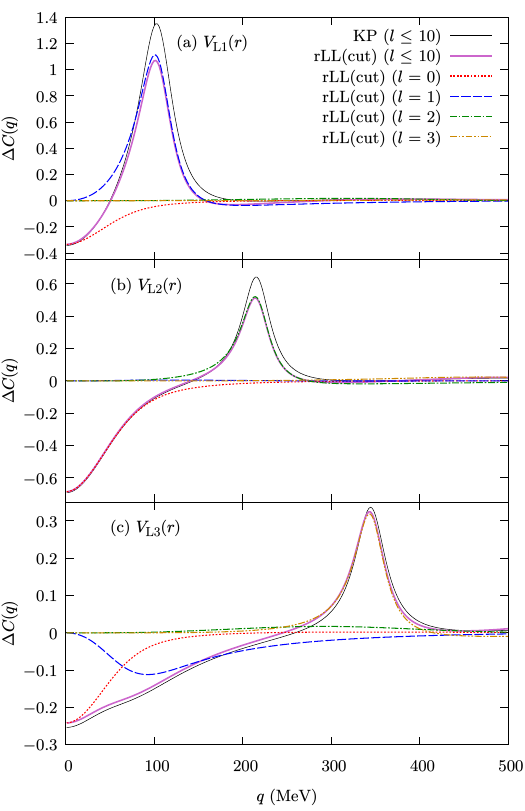}
  \caption{Corrections to the correlation function $\Delta C(q)$ by the
  regularized LL formula in the presence of a resonance. Panels~(a)--(c)
  correspond to the potentials $V_\mathrm{L1}(r)$, $V_\mathrm{L2}(r)$, and
  $V_\mathrm{L3}(r)$, respectively. The solid lines show the sum of the
  corrections to the correlation functions of $0 \le l \le 10$ based on the
  KP formula (think black line) and \rLLc{} (magenta line).  The red dashed,
  blue dotted, green dash-dotted, and yellow dash-dot-dotted lines show the
  partial-wave contributions of $l=1$, $\ldots$, $4$, respectively.  The
  reduced mass is $\mu = 469\MeV$, the source size is $R=2\fm$, and the cutoff
  parameter is $r_c = 1.0\fm$.  The vertical solid black lines show the real
  part of the pole position in the partial waves of $l=1$, $2$, and $3$ for
  panels~(a)--(c), respectively.}
  \label{fig:reso-l}
\end{figure}

\begin{figure}[thb]
  \centering
  \includegraphics[width = 0.45\textwidth]{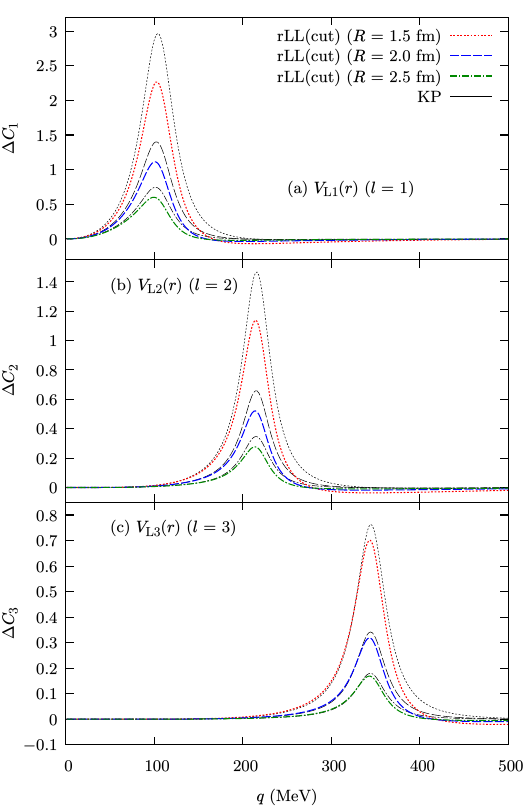}
  \caption{Source-size dependence of partial-wave contributions to the
  correlation function $\Delta C_l(q)$ by the regularized LL formula in the
  presence of a resonance.  Panels (a)--(c) correspond to the combinations of a
  potential and an angular momentum, $V_\mathrm{L1}(r)$ ($l=1$),
  $V_\mathrm{L2}(r)$ ($l=2$), and $V_\mathrm{L3}(r)$ ($l=3$), respectively.
  The red dashed, blue dotted, and green dash-dotted lines represent the
  results with the source size $R=1.5$, $2.0$, and $2.5\fm$.  The thin black
  lines are the KP formula with the source size $R$ being that of the same line
  type of \rLLc{}.  The reduced mass is $\mu = 469\MeV$, and the cutoff
  parameter is $r_c = 1.0\fm$.}
  \label{fig:reso-R}
\end{figure}

\begin{figure}[thb]
  \centering
  \includegraphics[width = 0.45\textwidth]{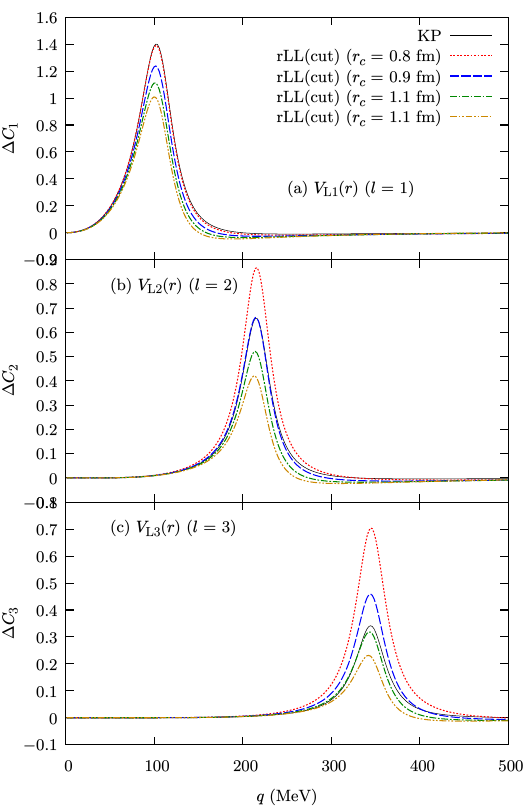}
  \caption{Cutoff dependence of partial-wave contributions to correlation
  functions in the presence of a resonance. The solid black line shows the
  exact integration of the KP formula.  The red dashed, blue dotted, green
  dash-dotted, and yellow dash-dot-dotted lines show the partial-wave
  contributions with the cutoff $r_c = 0.8$, $0.9$, $1.0$, and $1.1\fm$,
  respectively.}
  \label{fig:reso-c}
\end{figure}


\begin{figure}[thb]
  \includegraphics[width = 0.4\textwidth]{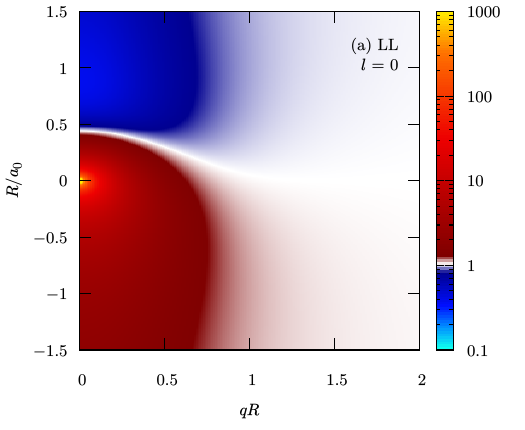}\\
  \includegraphics[width = 0.4\textwidth]{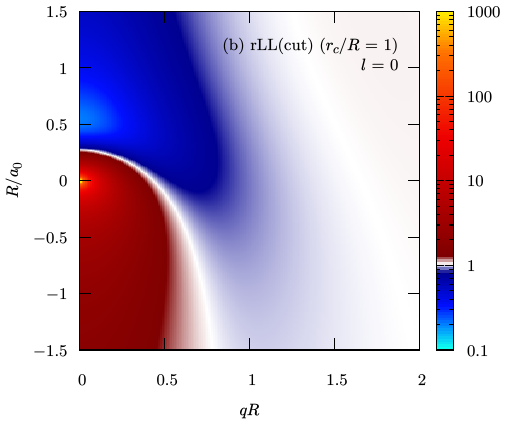}%
  \caption{Correlation with the $s$-wave contribution $C(q) = 1 + \Delta
  C_0(q)$ in the $(qR, R/a_0)$ plane. Panel (a) shows the result from the
  traditional LL formula, and panel (b) shows the result from \rLLc{}.  The
  $s$-wave scattering amplitude $f_0(q)$ is given by the effective-range
  expansion with the vanishing effective range, $r_0 = 0$.  The red area
  represents the region where the correlation function is increased by the $s$
  wave contribution, and the blue area represents the region where the
  correlation function is decreased.}
  \label{fig:ohnishi0}
\end{figure}

\begin{figure}[thb]
  \centering
  \includegraphics[width = 0.4\textwidth]{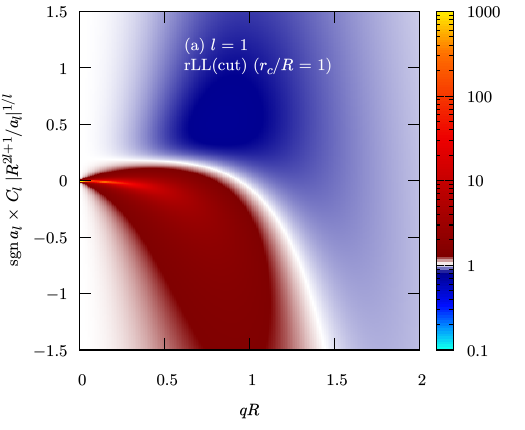} \\
  \includegraphics[width = 0.4\textwidth]{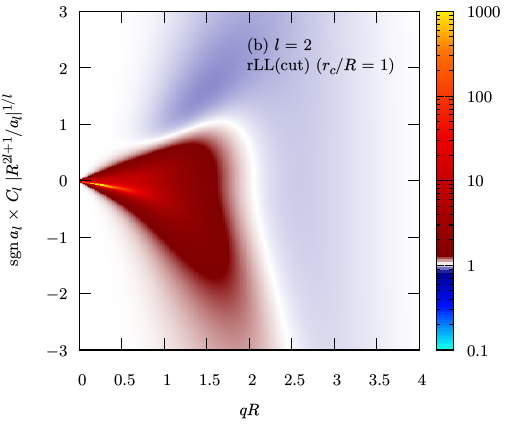} \\
  \includegraphics[width = 0.4\textwidth]{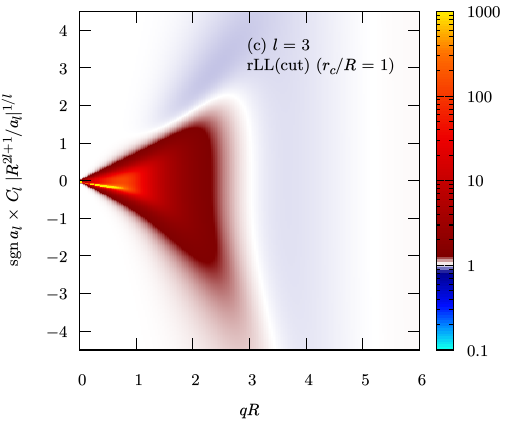}
  \caption{Correlation with a partial-wave contribution $C(q) = 1 + \Delta
  C_0(q)$ in the $(qR, (D_l |R^{2l+1} / a_l|^{1/l}) \sgn a_l)$ plane, where
  $D_l = 1/[(2l-1)!!(2l+1)!!]^{1/2l}$.  Panels (a)--(c) show the results for
  $l=1$, $2$, and $3$ with \rLLc{}, with $f_l$ given by the effective-range
  expansion. The effective-range parameter is assumed to be $r_l/R^{1-2l} =
  -1$.}
  \label{fig:ohnishi}
\end{figure}

\begin{figure}[thb]
  \centering
  \includegraphics[width = 0.4\textwidth]{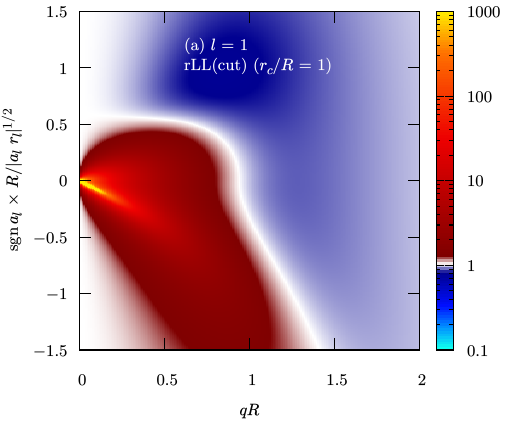} \\
  \includegraphics[width = 0.4\textwidth]{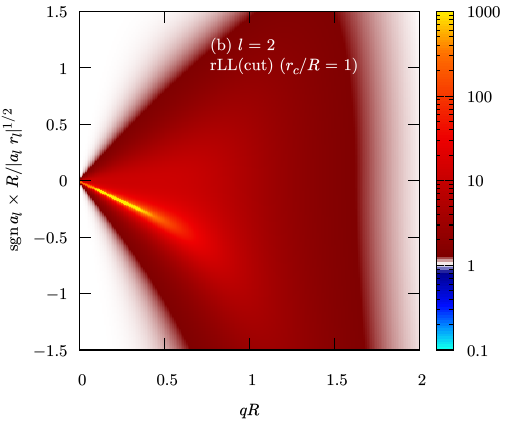} \\
  \includegraphics[width = 0.4\textwidth]{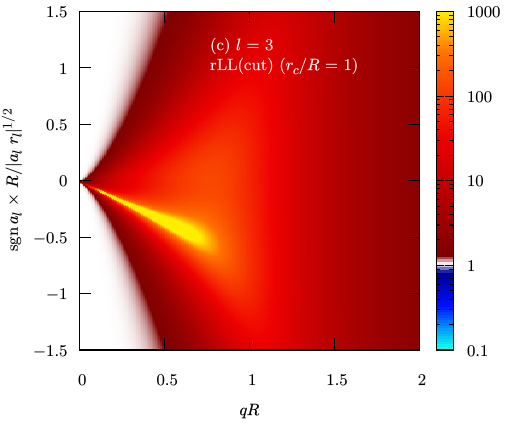}
  \caption{Correlation with a partial-wave contribution $C(q) = 1 + \Delta
  C_l(q)$ in the $(qR, (R/|a_l r_l|^{1/2})\sgn a_l)$ plane.  The same as
  Fig.~\ref{fig:ohnishi} except for the power of the vertical axis and the
  plotted region.  The cyan dotted lines show the linear
  relation~\eqref{eq:scale-resonance} for the approximate position of the
  resonance pole.}
  \label{fig:ohnishi2}
\end{figure}


In the previous section, we checked the behaviors of the regularized LL
formulae using the three potentials $V_\mathrm{2G}(r)$, $V_\mathrm{1GA}(r)$,
and $V_\mathrm{1GR}(r)$.  Using these potentials, we studied typical cases
without significant resonances.  We observed broad peak-like structures for
higher partial waves ($l\ge1$), but they are not associated with resonances,
and their magnitudes are usually very small compared to the $s$-wave
contribution.  However, the contributions from higher partial waves become
significant and important in the experimental data in the presence of the
resonances.

In this section, we prepare three one-range Gaussian potentials,
$V_\mathrm{L1}(r)$, $V_\mathrm{L2}(r)$, and $V_\mathrm{L3}(r)$, and adjust
their depths so that a resonance appears in the partial waves, $l=1$, $2$, and
$3$, respectively.  The potential form is given in the same form~\eqref{def:V}
as in the previous section, but with new labels, $\alpha = \mathrm{L1}$,
$\mathrm{L2}$, and $\mathrm{L3}$.  The parameters of the potentials are
summarized in Table~\ref{tab:reso-potparams}\@.  The positions of the poles of
$f_l(q)$ for $V_\mathrm{L1}(r)$, $V_\mathrm{L2}(r)$, and $V_\mathrm{L3}(r)$ are
$q = 105.22 -23.483i\MeV$ ($l=1$), $216.27 -19.776i\MeV$ ($l=2$), and $345.22
-21.186i\MeV$ ($l=3$), respectively.

Figure~\ref{fig:reso-l} shows partial-wave contributions of the correlation
function for the three potentials $V_\mathrm{L1}(r)$, $V_\mathrm{L2}(r)$, and
$V_\mathrm{L3}(r)$.  The solid lines show the sum of the partial-wave
contributions of $0 \le l \le 10$.  The thin solid black lines show the KP
formula, and the bold solid purple lines show \rLLc{}.  We can see that \rLLc{}
reproduces the qualitative behavior of the KP formula, though there are
quantitatively 0--20\% differences between them.  The red dashed, blue dotted,
green dash-dotted, and yellow dash-dot-dotted lines show the partial-wave
contributions from $l=0,\dots,3$, respectively.  We find that the partial wave
with the presence of the resonance [i.e., $l=1$ for $V_\mathrm{L1}(r)$, $l=2$
for $V_\mathrm{L2}(r)$, and $l=3$ for $V_\mathrm{L3}(r)$] has a dominant
contribution to the correlation function around the resonance position, and the
resonance positions are consistent with the real part of the corresponding pole
positions.  The magnitudes of the peaks are of order unity, which means that
the arguments made with Fig.~\ref{fig:kp-higher} that the magnitudes of the
contributions of higher partial waves does not hold in the presence of a
resonance.

Figure~\ref{fig:reso-R} shows the source-size dependence of the correlation
function in the partial wave with the resonance.  The thin and bold lines
correspond to the KP formula and \rLLc{}, respectively.  The dotted, dashed, and
dash-dotted lines correspond to the source size $R=1.5$, 2.0, and $2.5\fm$,
respectively.  Although there is a quantitative difference, both the KP formula
and \rLLc{} qualitatively behave in the same way.  The resonance peak is
sharper with a smaller source size $R$, and it becomes less significant with a
larger source size.

Figure~\ref{fig:reso-c} shows the cutoff dependence of the partial-wave
contributions of the correlation function.  The solid black lines show the KP
formula, and the other lines show \rLLc{} with different values of the cutoff
$r_c$.  We find that, with a proper choice of the cutoff parameter, the KP
formula can be quantitatively reproduced by \rLLc{}.  When the cutoff is
$r_c=0.8\fm$, the magnitude of the resonance peak of $V_\mathrm{L1}(r)$ in
$l=1$ is precisely reproduced. Similarly, $r_c=0.9\fm$ and $r_c=1.0\fm$
reproduce the magnitudes of the resonance peaks for $V_\mathrm{L2}(r)$ and
$V_\mathrm{L3}(r)$, respectively.  Those values of $r_c=0.8$--$1.0\fm$ are
slightly smaller than the Gaussian range $B_{L1,1}^{-1/2} = B_{L2,1}^{-1/2} =
B_{L3,1}^{-1/2} =1.25\fm$ but are of the same order.  The effective-range
parameters $r_l$ are $r_1 = -2.27032\fm^{-1} = (-0.440467\fm)^{-1}$ for
$V_\mathrm{L1}(r)$, $r_2 = -3.70273\fm^{-3} = (-0.646387\fm)^{-3}$ for
$V_\mathrm{L2}(r)$, and $r_3 = 28.3375 \fm^{-5} = (0.512304\fm)^{-5}$ for
$V_\mathrm{L3}(r)$, which have smaller length scales than the Gaussian range.

\section{Interaction-dependence of correlation functions}
\label{sec:ohnishi}

To understand the behavior of the correlation function for the $s$-wave
interaction intuitively, it has been useful to visualize the correlation
function as a heatmap in two dimensions $(qR, R/a_0)$~\cite{Ohnishi:2016elb,
Kamiya:2021hdb} based on the LL formula for the $s$ wave.  The horizontal axis
presents the $q$-dependence of the correlation function.  The vertical axis
shows the inverse of the scattering length, $1/a_0$, which corresponds to the
nature of the interaction, such as the repulsive or attractive nature, or the
presence of shallow bound states.  We now have a regularized LL formula for
higher partial waves, so it is insightful to visualize the correlation function
in a similar way.

Let us first compare the behavior of the traditional LL formula and the
regularized LL formula, \rLLc{}, in the $s$-wave case.
Figure~\ref{fig:ohnishi0} shows the correlation function with only the $s$-wave
contribution, $C(q)=1+\Delta C_0(q)$.  Panels (a) and (b) show the results with
the traditional LL formula and \rLLc{}, respectively.  The effective range
$r_0$ is assumed to vanish, i.e., $f_0(q) = 1/(-1/a_0 - iq)$, and thus the
effective-range correction in the LL formula also vanishes.  The cutoff of
\rLLc{} is assumed to be $r_c/R = 1$.

The red and blue areas show the regions where the $s$-wave contribution to the
correlation function is positive and negative, respectively.  In the
traditional LL formula, the $s$-wave contribution for negative $1/a_0$ is
positive in the entire $q$ region, but we find that a negative pocket arises
with the regularization in \rLLc{}.
This negative pocket is related to the lower bound of the integral domain in
the spherical-source KP formula~\eqref{eq:sphKP}, which gives the largest
weight $S(r)$ to the wave function $r^2|R_0(r)|^2$.  A small phase shift
$\delta_0$ contributes to the wave function as
\begin{align}
  r^2 |R_0(r)|^2
  &= \sin^2(qr+\delta_0) \notag \\
  &\approx \sin^2(qr)
    + \sin(2qr)\delta_0 + \mathcal{O}(\delta_0^2).
\end{align}
Because of the coefficient $\sin(2qr)$ in front of $\delta_0$, in a slightly
attractive case ($0 < \delta_0 \; (\approx -a_0 q) \ll 1$, i.e., $1/a_0 \ll
-q$), $r^2|R_0(r)|^2$ increases in the intervals $qr \in (n\pi, n\pi + \pi/2)$
and decreases in the intervals $qr \in (n\pi + \pi/2, (n+1)\pi)$, where $n$ is
a nonnegative integer.  In the regularized LL formula, because of the
monotonically decreasing source function $S(r)$, the dominant contribution
comes from the change of the wave function near the lower bound of the integral
domain, from $r = r_c$ to $r \sim 2r_c$.  As a result, the $s$-wave
contribution becomes positive and negative alternately as a function of the
horizontal axis $qR = (qr_c)(R/r_c)$.


We shall now check the behavior for higher partial waves.
Figure~\ref{fig:ohnishi} shows the correlation functions with only a single
partial-wave contribution of the angular momentum $l$, $C(q) = 1 + \Delta
C_l(q)$.  Each panel shows the result for a different $l$.  The effective-range
parameter $r_l$ in the unitary limit $1/a_l \to 0$ for higher partial waves
$(l\ge1)$ needs to be negative in three-dimensional space due to the limitation
stemming from Wigner's causal bound~\cite{Wigner:1955zz, Hammer:2009zh,
Hammer:2010fw}.  To cover a wide range of $1/a_l$ including the unitary limit,
we fix $r_l/R = -1$ for higher partial waves ($l\ge1$) in the present analysis.

The vertical axis of Fig.~\ref{fig:ohnishi} is chosen to be $D_l
|R^{2l+1}/a_l|^{1/l}\sgn a_l \propto 1/|a_l|^{1/l}$, where $D_l =
1/[(2l-1)!!(2l+1)!!]^{1/2l}$, because of the following scaling of the partial-wave
contributions for small $q$ and $1/a_l$: We first consider the scaling of $\hat
S_l$~\eqref{def:Sc}.  The asymptotic form of the Riccati--Hankel function $\hat
h_l(x)$ at small $x = qr$ is written as
\begin{align}
  \hat h_l(qr) \approx \frac{(qr)^{l+1}}{(2l+1)!!} - i\frac{(2l-1)!!}{(qr)^l}.
\end{align}
Therefore, at the lowest order in $q$, the real and imaginary parts of $\hat S_l$
are written as
\begin{align}
  \Re \hat S_l &\approx -\frac{[(2l-1)!!]^2}{q^{2l}} \langle r^{-2l} \rangle, \\
  \Im \hat S_l &\approx -\frac{2q}{2l+1} \langle r\rangle,
\end{align}
where $\langle r^m\rangle := \int_{r_c}^\infty S(r) r^m dr$ is the $m$th moment
of $S(r)$ with the cutoff $r_c$.  Similarly, the lowest orders of the real and
imaginary parts of the effective range expansion~\eqref{eq:ERE} are written as
\begin{align}
  \Re f_l &\approx -a_lq^{2l}, \\
  \Im f_l &\approx a_l^2 q^{4l+1}.
\end{align}
Thus, \rLLc{}~\eqref{eq:correction.hatS} is evaluated at the lowest order in
$q$ as
\begin{align}
  \Delta C_l(q) &\approx
    [(2l+1)!!(2l-1)!! a_l^2 \langle r^{-2l} \rangle
    -2 a_l \langle r\rangle] q^{2l}.
\end{align}
When $|R^{2l+1}/a_l|$ is small, the first term dominates:
\begin{align}
  \Delta C_l(q) &\approx
    \langle r^{-2l} \rangle \biggl(\frac{q}{D_l/|a_l|^{1/l}}\biggr)^{2l} \notag \\
    &= \biggl(\frac{\langle r^{-2l} \rangle}{R^{-2l-2}}\biggr)
    \biggl[\frac{qR}{D_l|R^{2l+1}/a_l|^{1/l}}\biggr]^{2l}.
  \label{eq:qscale}
\end{align}
In the second line, quantities $\langle r^{-2l}\rangle$, $q$, and $a_l$ are
written in dimensionless combinations using $R$.  Here, $qR$ and
$D_l|R^{2l+1}/a_l|^{1/l}$ appear in the combination
$qR/D_l|R^{2l+1}/a_l|^{1/l}$.  This means that the contour lines near the
origin in Fig.~\ref{fig:ohnishi} are expected to become linear,
$D_l|R^{2l+1}/a_l|^{1/l} \propto qR$\@. In fact, we see the linear behavior of
the edges of the red areas in Fig.~\ref{fig:ohnishi}.

In Fig.~\ref{fig:ohnishi}, we find that the partial-wave contribution
vanishes at $qR=0$, except for $1/a_0=0$, because of the
scaling~\eqref{eq:qscale}, $\Delta C_l \propto (qR)^{2l}$.  This corresponds to
the fact that the wave function $R_l(r) \propto (qr)^l$ is suppressed near the
origin $r=0$ due to the repulsion by the centrifugal force.

We also observe a structure similar to the $s$-wave case. In particular, we
have a shallow negative pocket at larger $q$ in the $1/a_0 < 0$ region.  This
behavior of \rLLc{} is also observed in Fig.~\ref{fig:rLL-compare-R} for
\rLLc{} with the effective range expansion, but it is not present for \rLLc{}
with the exact $f_l(q)$ in Fig.~\ref{fig:rLL-compare-R}.  This means that a
simple argument based on the effective range expansion does not apply to a
large $q$ region in a realistic setup.


In Fig.~\ref{fig:ohnishi}, we also notice a sharp ridge near the origin in the
$1/a_l < 0$ region.  Figure~\ref{fig:ohnishi2} shows the behavior of the
same correlation functions near the origin with a different power in the
vertical axis, $(R/|r_la_l|^{1/2})\sgn a_l \propto |a_l|^{-1/2}$, where we can
see the ridge more clearly.  This ridge corresponds to the resonance in the
effective range expansion of $f_l(q)$~\eqref{eq:ERE}.  The pole of $f_l(q)$ is
determined at the lowest order of $q$ as
\begin{align}
  q \approx \biggl(\frac2{a_l r_l}\biggr)^{1/2},
  \quad (a_l r_l \ge 0),
  \label{eq:qscale-asqinv}
\end{align}
by solving
\begin{align}
  q^{2l} f_l^{-1} \approx -\frac1{a_l} + \frac{r_lq^2}2 -iq^{2l+1},
  \quad (l \ge 1),
  \label{eq:finv}
\end{align}
where we ignored the term $-iq^{2l+1} = \mathcal{O}(q^3)$, which has a higher
order in $q$ for higher partial waves ($l\ge1$).  Since the effective-range
parameter is expected to be negative $r_l < 0$ for higher partial waves
($l\ge1$) with small $1/a_l$, Eq.~\eqref{eq:qscale-asqinv} indicates that a
shallow bound state accompanied by a virtual state emerges for $1/a_l>0$,
whereas a pair of resonance and anti-resonance poles emerges for $1/a_l<0$.
Thus, the
resonance ridge is expected to appear in the $1/a_l < 0$ region, with the
linear shape,
\begin{align}
  qR \approx \sqrt{2}\frac{R}{|a_l r_l|^{1/2}}, \qquad (1/a_l < 0).
  \label{eq:scale-resonance}
\end{align}

We observe that the resonance ridge becomes broader for a larger $qR$, while
retaining the linear relation~\eqref{eq:scale-resonance}.  This implies that
the resonance peak in the correlation function is broadened as the source size
$R$ becomes larger, while the peak position $q\approx \sqrt{2}/|a_lr_l|^{1/l}$
is unchanged.  We also find that the resonance ridge survives in a larger $q$
region for a larger angular momentum $l$.  This suggests that one may look at
larger source sizes to find the resonances of higher partial waves.

\section{Conclusion}
\label{sec:conclusion}

In recent years, femtoscopy has increasingly been recognized as an important
tool for constraining hadron--hadron interactions, in particular between exotic
hadrons, for which traditional scattering experiments are limited.
Although the Koonin--Pratt (KP)~\cite{Koonin:1977fh, Pratt:1984su,
Bauer:1992ffu} and Lednicky--Lyuboshitz (LL)~\cite{Lednicky:1981su} formulae
for the $s$-wave interaction are widely used to constrain the source function
and the interaction by the experimentally measured correlation function $C(q)$,
the real data are contaminated with contributions from higher partial waves in
general.  In our previous work~\cite{Murase:2024ssm}, we generalized the
existing spherical-source KP~\cite{Morita:2014kza} formula and the LL formula
to include the contributions from higher partial waves.  However, we also found
that the generalized LL formula suffers from an unphysical behavior of the
asymptotic function adopted by the formula.  In this study, we have attempted
to resolve the problem by introducing a regularization to the generalized LL
formula.

In Sec.~\ref{sec:rLL}, we have considered three regularizations using different
forms of the asymptotic wave functions and derived three regularized LL
formulae, \rLLc{} [Eqs.~\eqref{eq:correction.hatS} and~\eqref{def:Sc}], \rLLd{}
[Eqs.~\eqref{eq:correction.hatS} and~\eqref{def:Sd}], and \rLLw{}
[Eqs.~\eqref{eq:correction.hatSw} and~\eqref{def:Sw}], written in terms of the
source function $S(r)$ and the scattering amplitude $f_l(q)$.  We have then
compared the performance of the three regularized LL formulae in
Sec.~\ref{sec:rLL-result}.  Although the three regularized LL formulae produce
almost the same quality of results for different potentials in most cases, we
demonstrated that \rLLc{} produces stable results with a simple expression.  We
also investigated the source-size and cutoff-parameter dependencies of \rLLc{}
and the consequence of using the effective range expansion for $f_l(q)$.  We
then checked the behavior in the presence of resonances in
Sec.~\ref{sec:rLL-resonance}.  Finally, based on \rLLc{} with the effective
range expansion of the partial-wave scattering amplitude $f_l(q)$, we presented
heatmaps of the partial-wave contributions to the correlation functions,
$\Delta C_l$, as a function of the scattering-length parameter $1/a_l$ and the
momentum $q$ in Sec.~\ref{sec:ohnishi}.  We have shown the usefulness of this
type of plot in understanding the dependence of the correlation function on the
nature of the interaction and the source size.  In the heatmaps, we idenfied
scaling behaviors of the partial-wave contributions to the correlation
function---the scaling behavior~\eqref{eq:qscale} at small $q$, and the scaling
of the resonance peak~\eqref{eq:scale-resonance}---which will be useful in
understanding the general behavior of the correlation function for higher
partial waves.

\section*{Acknowledgement}
The authors thank Yuki Kamiya and Asanosuke Jinno for useful discussions.  The
work has been supported by JSPS KAKENHI Grant Numbers JP23H05439 (K.M. and
T.H.), JP23K13102 (K.M.), and JP22K03637 (T.H.).

\appendix
\section{Numerical methods}
\label{app:numerics}

We here describe the numerical methods to calculate the wave function $R_l(r)$
and related quantities for higher partial waves with a large $l$.  The
traditional way of the Numerov method has the fifth-order accuracy (with the
discretization error of $\Order(\Delta r^6)$ with $\Delta r$ being the step
size of the numerical integration) and produces very precise results for the
$s$-wave case.  For higher partial waves, the wave function at small $r$ is
expected to behave as $R_l \sim r^l$, so the step size needs to be very small
for $l \ge 6$ if one uses the Numerov method.  Even for $l \le 5$, it is useful
to implement a more robust and efficient integrator based on the embedded
Runge--Kutta method with the error control.  In this study, we transformed the
Schr\"odinger equation using variation of constants and numerically integrated
the equation for the Jost functions using an embedded Runge--Kutta method with
Hairer's adaptation of the Dormand--Prince pair with order 8(5,3)
(DOP853)~\cite{Hairer1993}.  We validated the results with the Numerov method
for lower $l$.  In this section, we describe the details of the method.

\subsection{Wave function and correlation function}

We first describe the method to calculate the wave function $R_l(r)$, the
scattering amplitude $f_l$, and the partial-wave contribution to the
correlation function $\Delta C_l$.  We here assume a complex momentum $q \in
\mathbb{C}$ because we also use the method in finding the poles of the $S$
matrix.  To solve the Schr\"odinger equation,
\begin{align}
  \biggl[
    -\frac{\partial^2}{\partial r^2}
    - \frac2r\frac{\partial}{\partial r}
    + \frac{l(l+1)}{r^2} - q^2 + v(r)\biggr] R_l(r) &= 0,
  \label{eq:app.schro}
\end{align}
with $v(r) := 2\mu V(r)$ being a short-range potential, it is useful to consider variation of constants with
the basis being the spherical Hankel functions, $h_l^{(1)}(qr) = j_l(qr) +
iy_l(qr)$ and $h_l^{(2)}(qr) = [h_l^{(1)}(q^*r)]^* = j_l(qr) - iy_l(qr)$.  The
solution in the free space, where $v(r) = 0$, can be represented by
\begin{align}
  R_l(r;q) &= c_1 h^{(1)}(qr) + c_2 h^{(2)}(qr),
\end{align}
where $c_1, c_2 \in \mathbb{C}$ are integral constants.  For variation of
constants, we replace $c_1$ and $c_2$ with undetermined functions $c_1(r;q)$
and $c_2(r;q)$:
\begin{align}
  R_l(r;q) &= c_1(r;q) h^{(1)}(qr) + c_2(r;q) h^{(2)}(qr).
\end{align}
Since we introduced two degrees of freedom $c_1(r;q)$ and $c_2(r;q)$, we have a
freedom of degree to specify a constraint:
\begin{align}
  \frac{\partial c_1(r;q)}{\partial r} h^{(1)}(qr) +
  \frac{\partial c_2(r;q)}{\partial r} h^{(2)}(qr) &= 0.
  \label{eq:app.constraint}
\end{align}
Combining the Schr\"odinger equation~\eqref{eq:app.schro} and the
constraint~\eqref{eq:app.constraint}, the evolution equation for the
coefficient functions $c_1(r)$ and $c_2(r)$ is obtained as
\begin{align}
  \frac{d}{dr}\begin{pmatrix} c_1(r) \\ c_2(r)
  \end{pmatrix} &= \frac{iqr^2v(r)R_l(r)}{2}\begin{pmatrix}
    -h^{(2)}(qr) \\ h^{(1)}(qr)
  \end{pmatrix}.
\end{align}
This can be augmented with the integration of the spherical-source KP
formula~\eqref{eq:sphKP-partial} as
\begin{align}
  \frac{d}{dr}\begin{pmatrix} c_1(r) \\ c_2(r) \\ I(r)
    \end{pmatrix} &= \begin{pmatrix}
      -\frac12 iqr^2v(r)R_l(r) h^{(2)}(qr) \\
      \frac12 iqr^2v(r)R_l(r) h^{(1)}(qr) \\
      4\pi r^2 S(r) [|R_l(r)|^2 - |j_l(qr)|^2]
  \end{pmatrix}.
\end{align}
We solve the above differential equation using the DOP853
integrator~\cite{Hairer1993}.  To normalize the solution at the origin as
$\varphi(r) \approx j_l(qr)$ ($r\to 0$), the initial conditions for the
coefficient functions are set to be $c_1(0) = c_2(0) = 1/2$.  The initial
condition for the KP integral $I(r)$ is $I(0) = 0$.

The coefficient functions $c_1(r)$ and $c_2(r)$ converge to the Jost function
at a sufficiently large $r$ where $v(r)$ vanishes:
\begin{align}
  c_2 := \lim_{r\to\infty} c_2(r)
  &= \frac12\biggl[1 + iq\int_0^\infty dr r^2 v(r)\varphi(r) h_l^{(1)}(r)\biggr] \notag \\
  &= \frac12 \Jost_l(q), \\
  c_1 := \lim_{r\to\infty} c_1(r)
  &= \frac12 \Jost_l(-q),
\end{align}
while the KP integral converges to the partial-wave contribution to the
correlation function at a sufficiently large distance $r\gg R$:
\begin{align}
  \Delta C_l(q) = (2l+1) \lim_{r\to\infty} I(r).
\end{align}
The phase shift is obtained by solving $e^{2i\delta_l(q)}
= S = c_1/c_2$:
\begin{align}
  \delta_l(q) = \frac1{2i}\ln\frac{c_1}{c_2}.
\end{align}
The partial-wave amplitude is
\begin{align}
  f_l(q) &= \frac{e^{2i\delta_l(q)} - 1}{2iq} = \frac{c_1/c_2 - 1}{2iq}.
\end{align}
The poles of $f_l(q)$ are obtained as the zeroes of $c_2(q)$.  For numerical
stability, we first obtain the zeroes $q\in\mathbb{C}$ of $c_1(q)$ and $c_2(q)$
by solving $\Re[c_2(q)c_1^*(q)] = \Im[c_2(q)c_1^*(q)] = 0$ using a nested
bisection method, and we then select the zeroes of $c_2(q)$ by examining the
values of $c_1(q)$ and $c_2(q)$.

\subsection{Effective-range parameters}

To calculate the effective-range parameters, $a_l$ and $r_l$, we consider the
small-$q$ expansion of the Schr\"odinger equation with respect to
$q$.  Unlike the method with naive numerical differentiation, this perturbative
method can be naturally extended for an arbitrary orders of the effective range
expansion robustly.

We first represent the wave function in the following form:
\begin{align}
  R_l(r) &= \frac{(qr)^l}{(2l+1)!!}[c_J(r) \tilde j_l(qr) + c_Y(r) \tilde y_l(qr)],
  \label{eq:effective-range-expansion.wf}
\end{align}
with $(c_J, c_Y)$ defined as the variable transform of $(c_1, c_2)$ as
\begin{align}
  c_1(r) + c_2(r) &= c_J(r), \\
  c_1(r) - c_2(r) &= c_Y(r) \frac{i(qr)^{2l+1}}{(2l-1)!!(2+1)!!}.
\end{align}
The functions $\tilde j_l(x)$ and $\tilde y_l(x)$ are defined to be the ratio
of $j_l(x)$ and $y_l(x)$ to the dominant term of each at small $x$:
\begin{align}
  j_l(x)
    &= \frac{x^l}{(2l+1)!!}
    \sum_{m=0}^\infty \frac1{m!}\frac{(-x^2/2)^m}{\prod_{k=1}^m(2l+1+2k)} \notag \\
    &= \frac{x^l}{(2l+1)!!} \tilde j_l(x), \\
  y_l(x)
    &= -\frac{(2l-1)!!}{x^{l+1}}
    \sum_{m=0}^\infty \frac1{m!}\frac{(x^2/2)^m}{\prod_{k=1}^m(2l+1-2k)} \notag \\
    &= -\frac{(2l-1)!!}{x^{l+1}} \tilde y_l(x).
\end{align}
Both $\tilde j_l(x)$ and $\tilde y_l(x)$ are the functions of $x^2$ and have
the form $1+\Order(x^2)$.

The evolution equation for $c_J(r)$ and $c_Y(r)$ is
\begin{align}
  \frac{d}{dr}\begin{pmatrix} c_J(r) \\ c_Y(r) \end{pmatrix}
  &= \frac{rv(r)\tilde R_l(r)}{2l+1}
    \begin{pmatrix} \tilde y_l(qr) \\ -\tilde j_l(qr) \end{pmatrix}
    - \begin{pmatrix} 0 \\ \frac{2l+1}r c_Y(r) \end{pmatrix},
  \label{eq:effective-range-expansion.E-eom}
\end{align}
where
\begin{align}
  \tilde R_l(r)
    &:= c_J(r) \tilde j_l(qr) + c_Y(r) \tilde y_l(qr).
\end{align}
The asymptotic behavior of $c_J(r)$ and $c_Y(r)$ at small $r$ is constrained by
the initial conditions for $c_1(r)$ and $c_2(r)$ and the regularity of the wave
function~\eqref{eq:effective-range-expansion.wf}:
\begin{align}
  c_J(r) &= 1 + \Order(r^\alpha), \quad
  c_Y(r) = \Order(r^\beta), \quad (r\to 0),
  \label{eq:effective-range-expansion.E-asymp-1}
\end{align}
where $\alpha > 0$ and $\beta \ge -l$.  The equations can be solved up to a
sufficiently large $r = R$ such that $v(r) = 0$ ($r > R$).  For $r>R$, the
equation simplifies as
\begin{align}
  \frac{d}{dr}\begin{pmatrix} c_J \\ c_Y \end{pmatrix}
  &= \begin{pmatrix} 0 \\ -\frac{2l+1}{r}c_Y \end{pmatrix}, \quad \text{($r > R$)},
\end{align}
and the solution at $r > R$ is given by
\begin{align}
  \begin{pmatrix} c_J(r) \\ c_Y(r) \end{pmatrix}
  &= \begin{pmatrix} c_J(R) \\ \bigl(\frac{R}{r}\bigr)^{2l+1} c_Y(R) \end{pmatrix}, \quad \text{($r > R$)}.
  \label{eq:effective-range-expansion.E-eval-at-R}
\end{align}

We notice that the $q$ dependence of
Eq.~\eqref{eq:effective-range-expansion.E-eom} enters only in $\tilde j_l$ and
$\tilde y_l$ in the combination of $(qr)^2$.  We consider obtaining the
solution for $c_Y(r)$ and $c_Y(r)$ in a perturbation with respect to $q^2$:
\begin{align}
  c_J(r) &= c_J^{(0)}(r) + q^2 c_J^{(2)}(r) + \cdots, \\
  c_Y(r) &= c_Y^{(0)}(r) + q^2 c_Y^{(2)}(r) + \cdots.
\end{align}
The basis functions are also expanded as
\begin{align}
  \tilde j_l(qr) &= \tilde j_l^{(0)} + q^2 \tilde j_l^{(2)}(r) + \cdots, \\
  \tilde y_l(qr) &= \tilde y_l^{(0)} + q^2 \tilde y_l^{(2)}(r) + \cdots,
\end{align}
where
\begin{align}
  \tilde j_l^{(0)} &= 1, & \tilde j_l^{(2)}(r) &= -\frac{r^2}{2(2l+3)}, & \cdots, \\
  \tilde y_l^{(0)} &= 1, & \tilde y_l^{(2)}(r) &= \frac{r^2}{2(2l-1)}, & \cdots.
\end{align}

The equations for $\Order(q^0)$ and $\Order(q^2)$ read
\begin{align}
  &\frac{d}{dr}\begin{pmatrix} c_J^{(0)} \\ c_Y^{(0)} \end{pmatrix}
  = \frac{rv(r)}{2l+1}
    \begin{pmatrix} 1 \\ -1 \end{pmatrix}
    (c_J^{(0)} + c_Y^{(0)})
  - \frac{2l+1}r \begin{pmatrix} 0 \\ c_Y^{(0)} \end{pmatrix},
    \label{eq:schro.ere.0} \\
  &\frac{d}{dr}\begin{pmatrix} c_J^{(2)} \\ c_Y^{(2)} \end{pmatrix}
  = \frac{rv(r)}{2l+1}
    \begin{pmatrix} 1 \\ -1 \end{pmatrix}
    (c_J^{(2)} + c_Y^{(2)})
  - \frac{2l+1}r \begin{pmatrix} 0 \\ c_Y^{(2)} \end{pmatrix} \notag \\
  & \quad +\frac{rv(r)}{2l+1} \begin{pmatrix}
      \tilde j_l^{(2)} + \tilde y_l^{(2)} & 2\tilde y_l^{(2)} \\
      -2\tilde j_l^{(2)} & -\tilde j_l^{(2)} - \tilde y_l^{(2)}
    \end{pmatrix}
    \begin{pmatrix} c_J^{(0)} \\ c_Y^{(0)} \end{pmatrix}.
    \label{eq:schro.ere.2}
\end{align}
It should be noted that the coefficients in Eqs.~\eqref{eq:schro.ere.0}
and~\eqref{eq:schro.ere.2} are all real-valued.  We can show that
$c_J^{(2m)}(r)$ and $c_Y^{(2m)}(r)$ are real-valued in general for a
nonnegative integer $m$.
With the constraints~\eqref{eq:effective-range-expansion.E-asymp-1}, the
asymptotic form of the solution at small $r$ is obtained as
\begin{align}
  c_J^{(0)}(r)
    &\approx 1 + \int_0^r dr' \frac{r'v(r')}{2l+1} \notag \\
    &\approx 1 + \frac{1}{\gamma+1}\frac{r^2v}{2l+1}, \\
  c_Y^{(0)}(r)
    &\approx -\frac1{r^{2l+1}}\int_0^r dr' \frac{r'^{2l+2}v(r')}{2l+1} \notag \\
    &\approx -\frac1{\gamma+2l+2}\frac{r^2v}{2l+1},
\end{align}
where $\gamma$ is the power of the lowest order of $r$ in $rv$: $rv = r^\gamma
(1 + \Order(r))$. To ensure the existence of $c_Y^{(2)}$, $\gamma$ needs to
satisfy $\gamma = \beta - 1 \ge -l - 1$.  When the solution exists,
Eqs.~\eqref{eq:schro.ere.0} and~\eqref{eq:schro.ere.2} are integrated up to $r
= R$ with the initial conditions $c_J^{(0)} = 1$ or otherwise $c_J^{(2m)} =
c_Y^{(2m)} = 0$.  We use the DOP853 integrator~\cite{Hairer1993} to solve
Eqs.~\eqref{eq:schro.ere.0} and~\eqref{eq:schro.ere.2}.

The effective range expansion of the Jost function is obtained as
\begin{align}
  &\Jost_l(q) = 2c_2(\infty) \notag \\
  &\quad = c_J(\infty) - i\frac{(qr)^{2l+1}}{(2l-1)!!(2l+1)!!} c_Y(\infty) \notag \\
  &\quad = [c_J^{(0)}(R) + q^2 c_J^{(2)}(R) + \Order(q^4)] \notag \\
  &\qquad - i\frac{(qR)^{2l+1}}{(2l-1)!!(2l+1)!!} [c_Y^{(0)}(R) + q^2 c_Y^{(2)}(R) + \Order(q^4)] \notag \\
  &\quad =: c_J(R, q^2) - iq^{2l+1} \tilde c_Y(R, q^2),
\end{align}
where we used Eq.~\eqref{eq:effective-range-expansion.E-eval-at-R} to evaluate
$c_J(r)$ and $c_Y(r)$ at $R$.  The effective range expansion for the
partial-wave amplitude is obtained as
\begin{align}
  f_l &= \frac{\Jost_l(-q) - \Jost_l(q)}{2iq \Jost_l(q)} \notag \\
  &= \frac{2iq^{2l+1} \tilde c_Y(R)}{2iq[ c_J(R) - i q^{2l+1} \tilde c_Y(R)]} \notag \\
  &= \frac{q^{2l}}{c_J(R)/\tilde c_Y(R) - i q^{2l+1}},
\end{align}
where the nontrivial part is expanded as
\begin{align}
  \frac{c_J(R)}{\tilde c_Y(R)}
  &= \frac{(2l-1)!!(2l+1)!!}{R^{2l+1}}\frac{c_J(R)}{c_Y(R)} \notag \\
  &= -\frac1{a_l} + \frac12 r_l q^2 + \dots.
\end{align}
The effective-range parameters $a_l$ and $r_l$ are obtained as
\begin{align}
  a_l &= - \frac{R^{2l+1}}{(2l-1)!!(2l+1)!!}\frac{c_Y^{(0)}(R)}{c_J^{(0)}(R)}, \\
  r_l &= -\frac2{a_l}\biggl[1 + \frac{c_J^{(2)}(R)}{c_J^{(0)}(R)}  - \frac{c_Y^{(2)}(R)}{c_Y^{(0)}(R)}\biggr].
\end{align}

\bibliography{paper_higher}
\end{document}